\documentclass[12pt,a4paper]{article}
\usepackage{graphicx}
\usepackage{subeqnarray}
\begin{document}
%
%
%
%
\def\astrobj#1{#1}
\newenvironment{lefteqnarray}{\arraycolsep=0pt\begin{eqnarray}}
{\end{eqnarray}\protect\aftergroup\ignorespaces}
\newenvironment{lefteqnarray*}{\arraycolsep=0pt\begin{eqnarray*}}
{\end{eqnarray*}\protect\aftergroup\ignorespaces}
\newenvironment{leftsubeqnarray}{\arraycolsep=0pt\begin{subeqnarray}}
{\end{subeqnarray}\protect\aftergroup\ignorespaces}
\newcommand{\diff}{{\rm\,d}}
\newcommand{\pprime}{{\prime\prime}}
\newcommand{\sgn}{{\rm sgn}}
\newcommand{\szeta}{\mskip 3mu /\mskip-10mu \zeta}
\newcommand{\FC}{\mskip 0mu {\rm F}\mskip-10mu{\rm C}}
\newcommand{\appleq}{\stackrel{<}{\sim}}
\newcommand{\appgeq}{\stackrel{>}{\sim}}
\newcommand{\legr}{\stackrel{<}{>}}
\newcommand{\leqgr}{\stackrel{\le}{>}}
\newcommand{\grle}{\stackrel{>}{<}}
\newcommand{\geqle}{\stackrel{\ge}{<}}
\newcommand{\Int}{\mathop{\rm Int}\nolimits}
\newcommand{\Nint}{\mathop{\rm Nint}\nolimits}
\newcommand{\range}{{\rm -}}
\newcommand{\displayfrac}[2]{\frac{\displaystyle #1}{\displaystyle #2}}
%
%
\title{
O and Fe abundance correlations and distributions
inferred for the  thick and thin disk}
\author{{$\qquad~$R.~Caimmi}\footnote{
{\it Physics and Astronomy Department, Padua Univ., Vicolo Osservatorio 3/2,
I-35122 Padova, Italy}
email: roberto.caimmi@unipd.it~~~
fax: 39-049-8278212}
\phantom{agga}}
%
%
\maketitle
\begin{quotation}
\section*{}
\begin{Large}
\begin{center}
Abstract

\end{center}
\end{Large}
\begin{small}

\noindent\noindent

A linear [Fe/H]-[O/H] relation is found for different stellar populations
in the Galaxy (halo, thick disk, thin disk) from a data sample obtained in
a recent investigation (Ram{\'\i}rez et al. 2013).   These correlations
support previous results inferred from poorer samples: stars display a
``main sequence'' expressed as [Fe/H] = $a$[O/H$]+b\mp\Delta b$
where a unit slope, $a=1$, implies a constant [O/Fe] abundance ratio.
Oxygen and iron empirical abundance distributions are then determined
for different subsamples, which are well explained by the theoretical
predictions of multistage closed-(box+reservoir) (MCBR) chemical evolution
models by taking into account the found correlations.   The interpretation
of these distributions in the framework of MCBR models gives us
clues about inflow/outflow rates in these different Galactic regions and their
corresponding evolution.   Outflow rate for the thick and the thin disks
are lower than the halo outflow rate.
Moreover if the thin disk built up
from the thick disk, both systems result of comparable masses.
Besides
that, the iron-to-oxygen yield ratio and the primary to not primary
contribution ratio
for the iron production are obtained from the data, resulting consistent
with SNII progenitor nucleosynthesis and with the iron production from
SNIa supernova events.

\noindent
{\it keywords -
Galaxy: evolution - Galaxy: formation - stars: evolution - stars: formation.}
\end{small}
\end{quotation}
\pagebreak

\section{Introduction} \label{intro}

Simple models of chemical evolution rely on some basic assumptions, namely
(1) universal stellar initial mass function (IMF), which implies identical
star generations both in space and in time, regardless of composition; (2)
instantaneous recycling, which
implies short-lived stars instantaneously die and long-lived stars exist
forever; (3) instantaneous mixing, which implies gas returned from dying stars
and gas inflowing from outside are instantaneously and uniformly mixed with
the pre-existing gas; (4) universal nucleosynthesis for assigned stellar mass, which
implies nuclide production is independent of the initial composition.   For
further details and application to the solar neighbourhood, an interested
reader is addressed to earlier investigations on disk (e.g., Pagel and
Patchett 1975) and halo (e.g., Ryan and Norris 1991) stars.

In reality, the above mentioned assumptions are valid only to a first extent.
More specifically, (1) leaving aside pop III stars, a top-heavy IMF seems to
be triggered by tidal interactions due to merger or accretion events (e.g.,
Rieke et al. 1980; Doane and Mathews 1993; Lopez-Sanchez 2010)
while, on the other hand, the IMF lower end depends on the temperature,
pressure and composition of gas turned
into a star generation (e.g., Conroy et al. 2013; Bekki 2013); (2) gas
recycling after
star death is delayed for long-lived stars such as type Ia supernova (SNIa)
progenitors (e.g., Haywood 2001); (3) gas returned after star death or
inflowing
from outside is uniformly mixed with the interstellar medium after a finite
time, or restricted into cells with little interaction with the
surroundings (e.g., Wilmes and K\"oppen 1995; Wiersma et al. 2009); (4)
stellar nucleosynthesis
depends on the initial composition, in addition to the initial mass, for a
class of nuclides called secondary elements, in contrast with primary elements
for which the dependence on the initial composition may be neglected to a
first extent (e.g., Pagel and Tautvaisiene 1995).    In particular, it
holds for oxygen and iron created in massive stars or SNII events 
(e.g., Yates et al. 2013).   Stellar lifetime also
depends on the initial metal abundance, but the trend changes only slightly
and can be neglected to a first extent (e.g., Yates et al. 2013).

Accordingly, simple models of chemical evolution can be used as a 
zeroth order approximation and comparison with results from more refined
models could be useful.
Let nuclides produced
under the validity of assumptions (1)-(4) above be defined as simple primary
(sp) elements (Caimmi 2013a, hereafter quoted as C13a) and nuclides produced
otherwise be defined as non simple primary (np) elements.  Let processes
yielding sp and np elements be defined as sp and np processes, respectively.

In this scenario,
predictions may be compared with
observations from a selected star sample and discrepancies could
provide valuable informations on the history of the parent population.   In
particular, predicted relative mass abundances maintain constant,
$Z_{\rm Q_1}/Z_{\rm Q_2}=$ const, or $Z_{\rm Q}/Z=$ const, where Q$_1$,
Q$_2$, Q, are selected elements heavier than He, or metals, and
$Z=\sum Z_{\rm Q_i}$ where all metals are included.
%

In terms of number abundances normalized to solar values,
according to the standard spectroscopic notation%
\footnote{In general, [Q$_1$/Q$_2]=\log(N_{\rm Q_1}/N_{\rm Q_2})-
\log(N_{\rm Q_1}/N_{\rm Q_2})_\odot$, where $N_{\rm Q}$ is the number density
of the element, Q.},
the above relation translates into [Q$_1$/H] = [Q$_2$/H]+$b_{\rm Q_1}$,
where $b_{\rm Q_1}$ depends on $Z_{\rm Q_1}/Z_{\rm Q_2}$ and its
solar counterpart (e.g., C13a).  That is a relation which can be inferred
from sample stars where both [Q$_1$/H] and [Q$_2$/H] are known.

To this aim, oxygen appears as a natural choice of sp element.
In fact, it is mainly
synthesized within type II supernova (SNII) progenitors, which makes the above
assumptions (1)-(4) hold to an acceptable extent.   Accordingly, 
[Q/H] = [O/H]+$b_{\rm Q}$ provided Q is also sp element, which is a
straight line of unit slope and intercept, $b_{\rm Q}$, on the
$\{{\sf O}[{\rm O/H}][{\rm Q/H}]\}$ plane.
On the other hand, an empirical
[Q/H]-[O/H] relation may be obtained for any element (sp or np), when the
number of objects is high enough, and then compared with theoretical
predictions to determine the contribution of sp and np processes to the mass
abundance, $Z_{\rm Q_{sp}}$ and $Z_{\rm Q_{np}}$.

Classical simple models are closed-box (CB), allowing neither gas inflow nor
gas outflow, which implies total (gas+star) mass remains unchanged (e.g.,
Pagel and Patchett 1975).   The addition of a reservoir to the box opens the
possibility of both gas inflow (from the reservoir to the box) and gas outflow
(from the box to the reservoir), while the total mass remains unchanged
provided the reservoir is included in the balance.   In other words, the model
may be conceived as ``closed'' with regard to the box+reservoir and
``open'' with regard to the box alone, or shortly closed-(box+reservoir)
(CBR).   For further details, an interested reader is addressed to the parent
papers (Caimmi 2011a, 2012a).   The first example of CBR model is shown in a
classical paper where the predicted chemical evolution of the halo fits to the
data (Hartwick 1976).   Thought the box and the reservoir are not explicitly
mentioned therein, gas inhibited from star formation may safely be conceived
as outflowing from the box (the halo therein) to the reservoir (the disk
therein).

In general, CBR models maintain the standard assumptions mentioned above for
CB models, with the addition of (i) mass conservation within the
box+reservoir; (ii) gas outflow from the box into the reservoir or gas inflow
into the box from the reservoir (1) at a rate proportional to the star
formation rate, and (2) with composition proportional to its conterpart within
the box; (iii) absence of star formation within the reservoir.

Mass conservation within the box+reservoir can be expressed as (e.g., Caimmi
2011a):
\begin{equation}
\label{eq:mcbr}
(1+\kappa)\frac{\diff s}{\diff t}=-\frac{\diff\mu}{\diff t}~~;
\end{equation}
where $\kappa$ is the flow parameter (ratio of inflowing or outflowing gas
rate to locking up gas rate into long-lived stars and remnants), $s$ and $\mu$
are long-lived star and gas mass fraction.   More specifically, with respect
to CB models, Eq.\,(\ref{eq:mcbr}) discloses different
flow regimes as: outflow regime $(\kappa>0)$, where less gas is available for
star formation (Hartwick 1976); stagnation
regime $(\kappa=0)$, where CBR models reduce to CB models (Searle 1972;
Searle and Sargent 1972); moderate inflow regime
$(-1<\kappa<0)$, where a slightly larger amount of gas is available for star
formation and the gas mass fraction monotonically decreases in time
(Caimmi 2007); steady inflow regime
$(\kappa=-1)$, where a larger amount of gas is available for star formation
and the gas mass fraction remains unchanged in time (Caimmi 2011a, 2012a);
strong inflow regime $(\kappa<-1)$ where a substantially
larger amount of gas is available for star formation and the gas mass fraction
monotonically increases in time (Caimmi, 2011a, 2012a).

A main feature of CBR (and {\it a fortiori} CB) models is that the theoretical
differential abundance distribution (TDAD),
$\psi_{\rm Q}=\log[\diff N/(N\diff\phi_{\rm Q})]$, as a function of the
normalized abundance, $\phi_{\rm Q}=Z_{\rm Q}/(Z_{\rm Q})_\odot$, is close
to a straight line for the cases of interest,
$\psi_{\rm Q}=\alpha_{\rm Q}\phi_{\rm Q}+\beta_{\rm Q}$, where Q is a selected
primary element synthesised within SNII progenitors and $N$ the total number
of long-lived stars (Pagel 1989; Malinie et al. 1993; Rocha-Pinto and Maciel
1996; Caimmi 2011a, 2012a, and earlier references therein).    The normalized
abundance, in turn, may be expressed to an acceptable extent as (e.g., Pagel
and Patchett 1975; Hartwick 1979):
\begin{equation}
\label{eq:phimu}
\phi_{\rm Q}=-\frac1{1+\kappa}\frac{\hat p_{\rm Q}}{(Z_{\rm Q})_\odot}\ln\mu
~~;
\end{equation}
where $\hat p_{\rm Q}/(1+\kappa)$ is the effective yield.   Accordingly, a
linear (or nearly linear) fit to the empirical differential abundance
distribution (EDAD) allows comparison between observations from stellar
populations and predictions of CBR models.
More specifically, oxygen EDAD can be fitted by a broken line with several
segments, where each
segment can be interpreted within the framework of CBR models.   A
change in the slope implies a transition between adjacent evolutionary stages.

In other words, oxygen EDAD can be reproduced via a chain of CBR models where
the flow rate is discontinuous in each transition, or shortly multistage
closed-(box+reservoir) (MCBR) models.   In particular, early stages appear to
undergo strong inflow regime, middle stages approach steady inflow regime, and
late stages exhibit low inflow or outflow regime (Caimmi 2011a, 2012a).   The
special case of steady inflow regime, where the inflowing gas balances the
amount of pre-existing gas turned into stars, finds a counterpart in the
results of hydrodynamical simulations, where quasi equilibrium is attained
between inflowing gas, outflowing gas, gas turned into stars
(e.g., Finlator and Dav\'e 2008; Dav\'e et al. 2011a,b, 2012).

The main features of MCBR models (Caimmi 2011a, 2012a, 2013b, the last quoted
henceforth as C13b), for any assigned sp element, Q, may be summarized as
follows.
\begin{description}
\item[$\bullet$]
Simple MCBR models predict a linear TDAD,
$\psi_{\rm Q}=\alpha_{\rm Q}\phi_{\rm Q}+\beta_{\rm Q}$.
\item[$\bullet$]
The slope, $\alpha_{\rm Q}$, and the
intercept, $\beta_{\rm Q}$, of this TDAD depend on
the effective yield, $\hat p_{\rm Q}/(1+\kappa)$, on the
initial abundance, $(\phi_{\rm Q})_i$, and on the initial and final gas mass
fraction, $\mu_i$ and $\mu_f$.
\item[$\bullet$]
The TDAD slope relates to the flow regime as follows.   The steady state
inflow regime
$(\kappa=-1)$ implies null slope $(\alpha=0)$.   The strong inflow regime
$(\kappa<-1)$ implies positive slope $(\alpha>0)$.   The weak inflow regime
$(-1<\kappa<0)$, the stagnation regime $(\kappa=0)$, and the outflow regime
$(\kappa>0)$ imply negative slope $(\alpha<0)$.
\item[$\bullet$]
The abundance ratio between two sp elements 
remains fixed during the evolution and equals the yield ratio which, in turn,
is inversely proportional to the TDAD slope ratio,
$\hat p_{\rm Q_1}/\hat p_{\rm Q_2}=[(Z_{\rm Q_1})_\odot/(Z_{\rm Q_2})_\odot]
(\alpha_{\rm Q_2}/\alpha_{\rm Q_1})$.
\item[$\bullet$]
For two different populations, P$_1$, P$_2$, the TDAD slope ratio relates to
the flow regime, as
$(\alpha_{\rm Q})_{\rm P_1}/(\alpha_{\rm Q})_{\rm P_2}=(1+\kappa_{\rm P_1})/
(1+\kappa_{\rm P_2})$.
\item[$\bullet$]
Any choice of input parameters (initial gas mass fraction, $\mu_i$; initial
star mass fraction, $s_i$; initial inflowed/outflowed gas mass fraction,
$D_i$; initial and final element abundance normalized to the solar value,
$(\phi_{\rm Q})_i$, $(\phi_{\rm Q})_f$; TDAD slope and intercept,
$\alpha_{\rm Q}$, $\beta_{\rm Q}$; IMF, $\Phi(m)$;
true yield, $\hat p_{\rm Q}$) produce the 
output parameters (flow parameter, $\kappa$; final gas mass fraction, $\mu_f$;
final star mass fraction, $s_f$; final inflowed/outflowed gas mass fraction,
$D_f$) which can be taken as input parameters for the subsequent stage of
evolution.
\end{description}

To this respect, MCBR models make a further step with respect to standard
simple models, where a single stage is considered, in that a sequence of
evolutionary stages is described, where the inflow/outflow rate and related
parameters change passing from a previous stage to a subsequent one.   In this
context, the predicted TDAD must necessarily relate to sp elements, oxygen in
particular.

On the other hand, the EDAD is usually inferred from iron, which
is easier to be detected in stellar atmospheres.   But iron cannot be
considered as instantaneously recycled (i.e. sp element) in that a
substantial fraction is produced via SNIa events, whose progenitors are
low-mass stars $(m\appleq8m_\odot)$ belonging to binary systems where the
members are sufficiently close.

Then in absence of oxygen abundance data inferred from
observations, an empirical [Fe/H]-[O/H] relation (e.g., Carretta et al. 2000;
Israelian et al. 2001) has necessarily to be used for determining oxygen EDAD
from a selected star sample.

In earlier papers, a linear fit to the empirical [Fe/H]-[O/H] relation has
been aimed to different extents: (a) to infer oxygen EDAD from iron
EDAD in absence of rich samples where oxygen abundance is known, such as halo
stars (Caimmi 2011a); (b) to provide a rigorous star classification within
globular clusters and other halo and disk populations (C13a).   An
additional result, restricted to iron fraction synthesised within SNII
progenitors, is that the iron-to-oxygen yield ratio can be independently
expressed via either a linear [Fe/H]-[O/H] relation with a unit slope
regardless of iron and oxygen TDAD, or a linear iron and oxygen TDAD
regardless of [Fe/H]-[O/H] relation (C13b).

According to the above considerations, with regard to a selected element, Q,
and to an assigned star sample, both the inferred empirical [Q/H]-[O/H]
relation and the EDAD for Q and O provide valuable clues for understanding the
evolution of the parent population.   In most cases, Q = Fe, as iron can
easily be detected.

The current paper deals with a sample including halo,
thick disk and thin disk nearby stars, for which both oxygen and iron
abundances have been well determined (Ram{\'\i}rez et al. 2013, hereafter
quoted as Ra13) as extension of a previous investigation (Ram{\'\i}rez et al.
2007).
More specifically, rich subsamples are available with accurate abundance
values for disk populations, with the
addition of a poor incomplete halo sample (Ram{\'\i}rez et al. 2012).

By using this much better sample of stars with available data for oxygen and
iron, we
would like to check our previous finding and conclusions doing what follows:
\begin{description}
\item[(i)]
completely studying O-Fe correlations,
\item[(ii)] 
determining oxygen and iron EDAD,
\item[(iii)]
performing linear fits to oxygen and iron EDAD above and comparing with
related TDAD in the framework of MCBR models,
\item[(iv)]
checking that the iron EDAD obtained from oxygen EDAD and [Fe/H]-[O/H]
relation reproduces the actually measured iron EDAD,
\item[(v)]
determining sp and np iron contributions from [Fe/H]-[O/H] relation,
\item[(vi)]
estimating the ratio between yields for iron and oxygen,
\item[(vii)]
analysing the above results in the framework of MCBR models to estimate the
outflow and inflow rates.
\item[(viii)]
obtaining the thick disk evolution and comparing these results with other
conclusions inferred from age and star formation rates (Haywood et al. 2013;
Snaith et al. 2014).
\end{description}

Basic informations on the data together with regression line analysis are
provided in Section \ref{data}, where the
original sample is subsampled according to star population and star class.
The inferred EDAD for oxygen and iron are
shown in Section \ref{resu}.   The results are discussed in Section \ref{disc}.
The conclusion is drawn in Section \ref{conc}.

\section{Data and regression line analysis} \label{data}

The data are taken from a sample $(N=825)$ of solar neighbourhood FGK-type
dwarf
stars in the metallicity range, $-2.6<$ [Fe/H] $<0.5$, for which [O/H] has
been determined using high-quality spectra and a non-LTE analysis of the 777
nm OI triplet lines, while a standard spectroscopic approach has been followed
for the evaluation of [Fe/H], where values of FeII have been taken as
representative.  For further details, an interested reader is addressed to
Ra13 and earlier researches
(Ram{\'\i}rez et al. 2007, 2012).

Subsamples can be extracted from the parent sample (HD) according to different
populations and different classes: a star of the sample belongs to halo (HH),
thick disk (KD), or thin disk (ND), if $P_{\rm XY}>0.5$, where XY = HH, KD,
ND, and $P$ is a probability inferred from kinematics, or it is uncertain
between thick and thin disk (KN), when $P_{\rm KD}\le0.5$, $P_{\rm ND}\le0.5$,
$P_{\rm HH}\ll\min(P_{\rm KD}, P_{\rm ND})$, respectively.
It is worth noticing the above mentioned uncertainty criterion, which prevents
a disk star from being assigned to the thick or thin subsystem, is more
restrictive with respect to what assumed in Ra13.

According to Ra13, sample stars are classified as
cool high-metallicity dwarfs
$(T_{\rm eff}<5100\,{\rm K};\log g>4.4;$ [Fe/H]$>-0.1)$, giants
$(T_{\rm eff}<5500\,{\rm K};$\linebreak$\log g<4.0)$, outliers (relative to
the main disk [O/Fe] vs. [Fe/H] trend), which
shall be denoted throughout the text as c, g, o, respectively.   The remaining
stars are classified here as ``normal'' dwarfs and denoted as n.

The number of stars belonging to different subsamples, as given in Ra13, are
listed in Table \ref{t:samp}.   Subsamples related to a stellar population and
a selected class shall be denoted as XYz,
where XY defines the population, as shown below with the addition of XY = KN,
HD, and z defines the star class, z = n, c, g, o.
\begin{table}
\caption{Number of stars in the subamples from the original sample for
different populations (pop) as:
 ND - thin disk; KD - thick disk; KN - uncertain if
belonging to  thin or thick disk; HH - halo; HD - original sample
(HD = ND + KD + KN + HH);
and class (cl) as: n - normal
dwarfs; c - cool high-metallicity dwarfs; g - giants; o - outliers; t - total
(t = n + c + g + o).}
\label{t:samp}
\begin{center}
\begin{tabular}{crrrrr} \hline
\multicolumn{1}{l}{pop:} &
\multicolumn{1}{c}{ND} &
\multicolumn{1}{l}{KD} &
\multicolumn{1}{c}{KN} &
\multicolumn{1}{c}{HH} &
\multicolumn{1}{c}{HD} \\
cl  &     &     &   &    &    \\
\hline
n   & 513 & 215 & 6 & 41 & 775 \\
c   &  10 &   1 & 0 &  0 &  11 \\
g   &  14 &  18 & 0 &  3 &  35 \\
o   &   1 &   3 & 0 &  0 &   4 \\
t   & 538 & 237 & 6 & 44 & 825 \\
\hline
\end{tabular}                     
\end{center}                      
\end{table}                       
Though halo subsamples are largely incomplete, still they shall be considered
for inference of preliminary results and comparison with related disk
counterparts.

\begin{figure}[t]  
\begin{center}      
\includegraphics[scale=0.8]{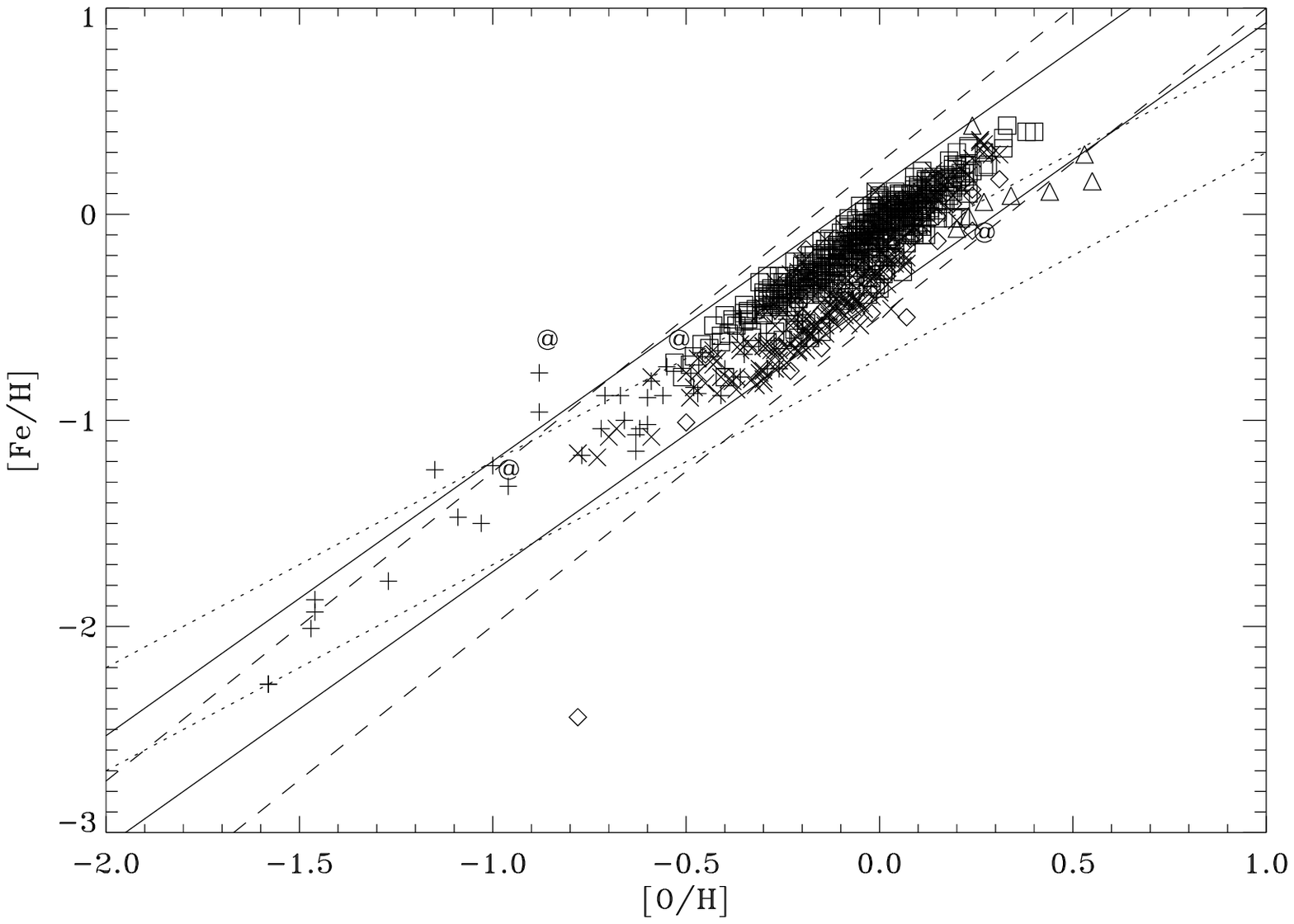}                      
\caption[ddbb]{The empirical [Fe/H]-[O/H] relation for HD stars
sampled in Ra13.   Caption of
symbols: normal main sequence dwarf stars - squares (NDn), saltires
(KDn), diagonalized squares (KNn), crosses (HHn);
cool main sequence dwarf stars - triangles (HDc); giants - diamonds (HDg);
outliers - ``at'' symbols (HDo).   Also shown for comparison are the ``main
sequences'', [Fe/H] = [O/H]$-$0.45$\mp$0.25 (dotted),
2[Fe/H] = 3[O/H]$-$0.25$\mp$0.75 (dashed), 3[Fe/H] = 4[O/H]$-$0.40$\mp$0.80
(full).   Typical error
bars are of the order of the symbol dimensions.   See text for further
details.}
\label{f:ofehk}     
\end{center}       
\end{figure}

The empirical [Fe/H]-[O/H] relation is plotted in Fig.\,\ref{f:ofehk} for 
the parent sample, HD, where different star classes are
denoted by different symbols, as squares (NDn), saltires (KDn), diagonalized
squares (KNn), crosses (HHn), triangles (HDc), diamonds (HDg), ``at'' (HDo).

Also shown are three selected ``main sequences'', expressed as:
\begin{equation}
\label{eq:FeO11}
[{\rm Fe/H}]=[{\rm O/H}]-0.45\mp0.25~~;
\end{equation}
(dotted), already shown in earlier researches (Caimmi 2012a; C13a);
\begin{equation}
\label{eq:FeO23}
2[{\rm Fe/H}]=3[{\rm O/H}]-0.25\mp0.75~~;
\end{equation}
(dashed), inferred from Ra13;
\begin{equation}
\label{eq:FeO34}
3[{\rm Fe/H}]=4[{\rm O/H}]-0.40\mp0.80~~;
\end{equation}
(full), inferred from Fig.\,\ref{f:ofehk}.
\begin{figure}[t]  
\begin{center}      
\includegraphics[scale=0.8]{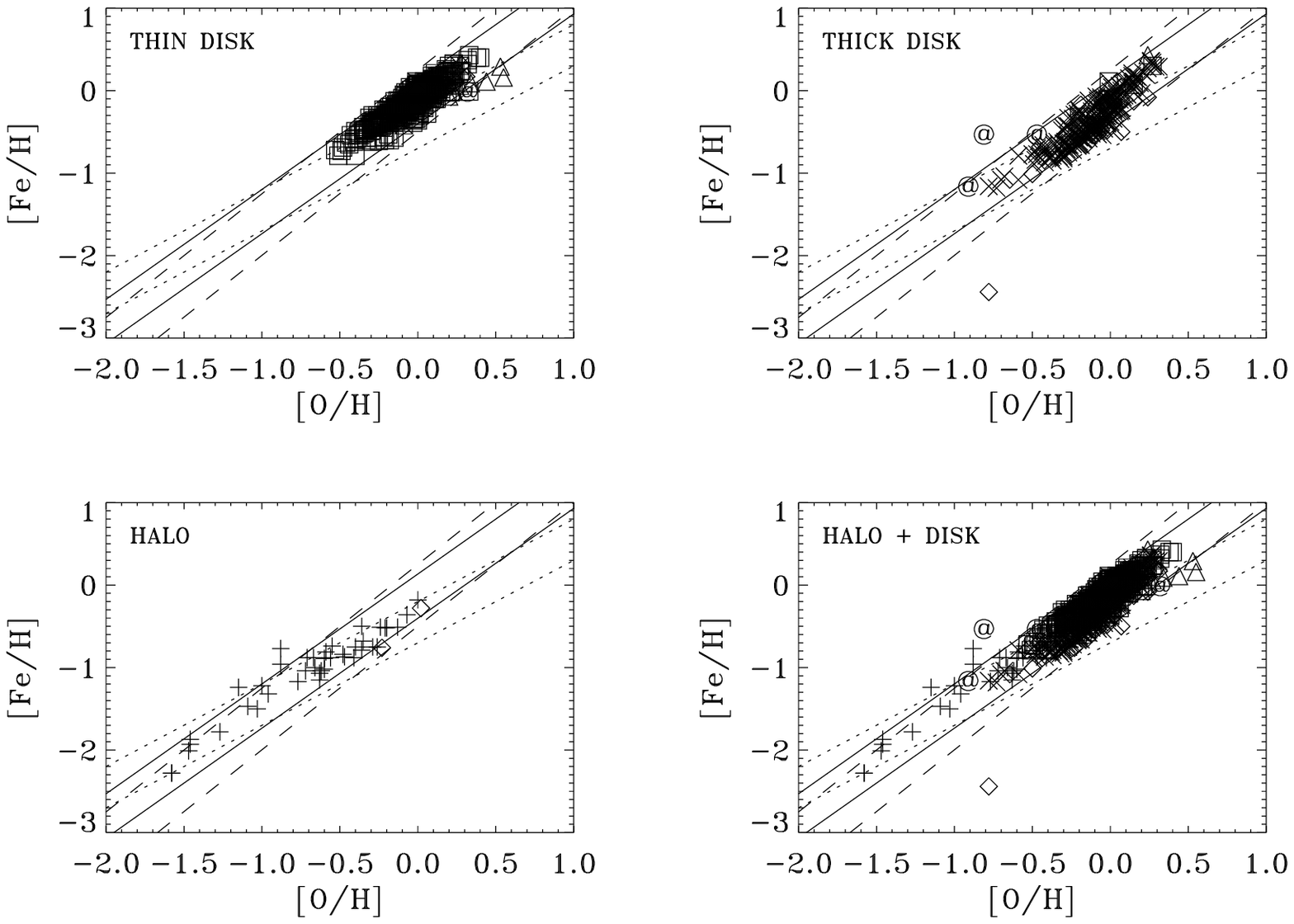}                      
\caption[ddbb]{The empirical [Fe/H]-[O/H] relation for ND (top left),
KD (top right), HH (bottom left), HD (bottom right) stars
sampled in Ra13.   Bottom right
panel is the repetition of Fig.\,\ref{f:ofehk} for better comparison.   Other
captions as in Fig.\,\ref{f:ofehk}.}
\label{f:ofehk4}     
\end{center}       
\end{figure}                                                                     

Keeping in mind typical errors are of the order of symbol dimensions on the
scale of the plot, an inspection of Fig.\,\ref{f:ofehk} shows a substantial
fraction of sample stars lie outside the main sequence of 
slope equal to 1.
Accordingly, iron cannot be considered as a simple primary element, contrary
to oxygen, mainly due to delayed recycling via SNIa events.

On the other hand, only a few stars lie outside the main sequences with slope
equal to 4/3 and 3/2, respectively, whose abundances and related uncertainties
are listed in Table \ref{t:outl}.   It includes stars from different
populations and classes: NDc (1), KDg (1), KDo (1), HHn (2), HHg (1), for a
total of 6.    In this view, the stars
under discussion should be considered as outliers.
\begin{table}
\caption{HIP number, oxygen and iron abundance, related uncertainty
($\Delta_{\rm Q}=\Delta$[Q/H], Q = O, Fe), population and
class (XYz), for stars lying outside (including the errors) the main sequences 
with slope equal to 4/3
and 3/2, respectively, plotted in Fig.\,\ref{f:ofehk}.   For further details 
refer to the text.}
\label{t:outl}
\begin{center}
\begin{tabular}{llllll} \hline
\multicolumn{1}{c}{HIP} &
\multicolumn{1}{c}{[O/H]} &
\multicolumn{1}{c}{$\Delta_{\rm O}$} &
\multicolumn{1}{c}{[Fe/H]} &
\multicolumn{1}{c}{$\Delta_{\rm Fe}$} &
\multicolumn{1}{c}{XYz} \\
\hline
069972 & $+$0.55 & 0.12 & $+$0.16 & 0.13 & NDc \\
060719 & $-$0.78 & 0.32 & $-$2.44 & 0.13 & KDg \\
049988 & $-$0.53 & 0.26 & $-$0.81 & 0.05 & KDo \\
012294 & $-$0.88 & 0.01 & $-$0.77 & 0.22 & HHn \\
068594 & $-$3.83 & 1.48 & $-$2.56 & 0.06 & HHg \\
114962 & $-$1.15 & 0.02 & $-$1.24 & 0.23 & HHn \\
\hline                            
\end{tabular}                     
\end{center}                      
\end{table}                       

In addition, the HHg star lies outside the scale of related
plots and cannot be represented even in the parent paper (Ra13).   If inferred
oxygen abundance is unbiased, the above mentioned star could be a globular 
cluster
outlier of second or later generation, which implies sodium overabundance 
(Ram{\'\i}rez et al. 2012).

The empirical [Fe/H]-[O/H] relation is plotted in Fig.\,\ref{f:ofehk4} for
ND, KD, HH subsamples and the parent sample, HD, the last repeated for better
comparison.
Within the current framework, all stars belonging to HH, KD, ND, DD = KD + ND
+ KN subsamples
could be used to infer both the [Fe/H]-[O/H] empirical relation and the EDAD
for O and Fe.   On the other hand, the empirical [O/Fe]-[Fe/H] relation is
inferred in Ra13 using stars of class n only, which are the main
part of sample stars as shown in Table \ref{t:samp}.   Accordingly, to be
consistent with the results of Ra13, further analysis shall be
restricted to stars of class n
and, to save space, subsamples HHn, KDn, NDn,
DDn, HDn, shall hereafter be denoted as HH, KD, ND, DD, HD, respectively.

At this stage, two considerations can be performed.   First, the sample
consists of nearby stars, which implies that our conclusions rely on the
assumption that the solar neighbourhood is a typical region of the related
subsystem (ND, KD, HH).   Second, although a bilinear
regression with a knee has been widely used for fitting the [O/Fe]-[Fe/H]
empirical relation (e.g., Ra13), we prefer a single linear regression.
The dichotomy between bilinear (e.g., Carretta et
al. 2000; Gratton et al. 2000) and linear (e.g., Israelian et al. 2001;
Takada-Hidei et al. 2001) trend is long-dating, but  the last
alternative is preferred here, as in previous works (Caimmi 2012a; C13a), for
reasons of simplicity.

\begin{figure}[t]  
\begin{center}      
\includegraphics[scale=0.8]{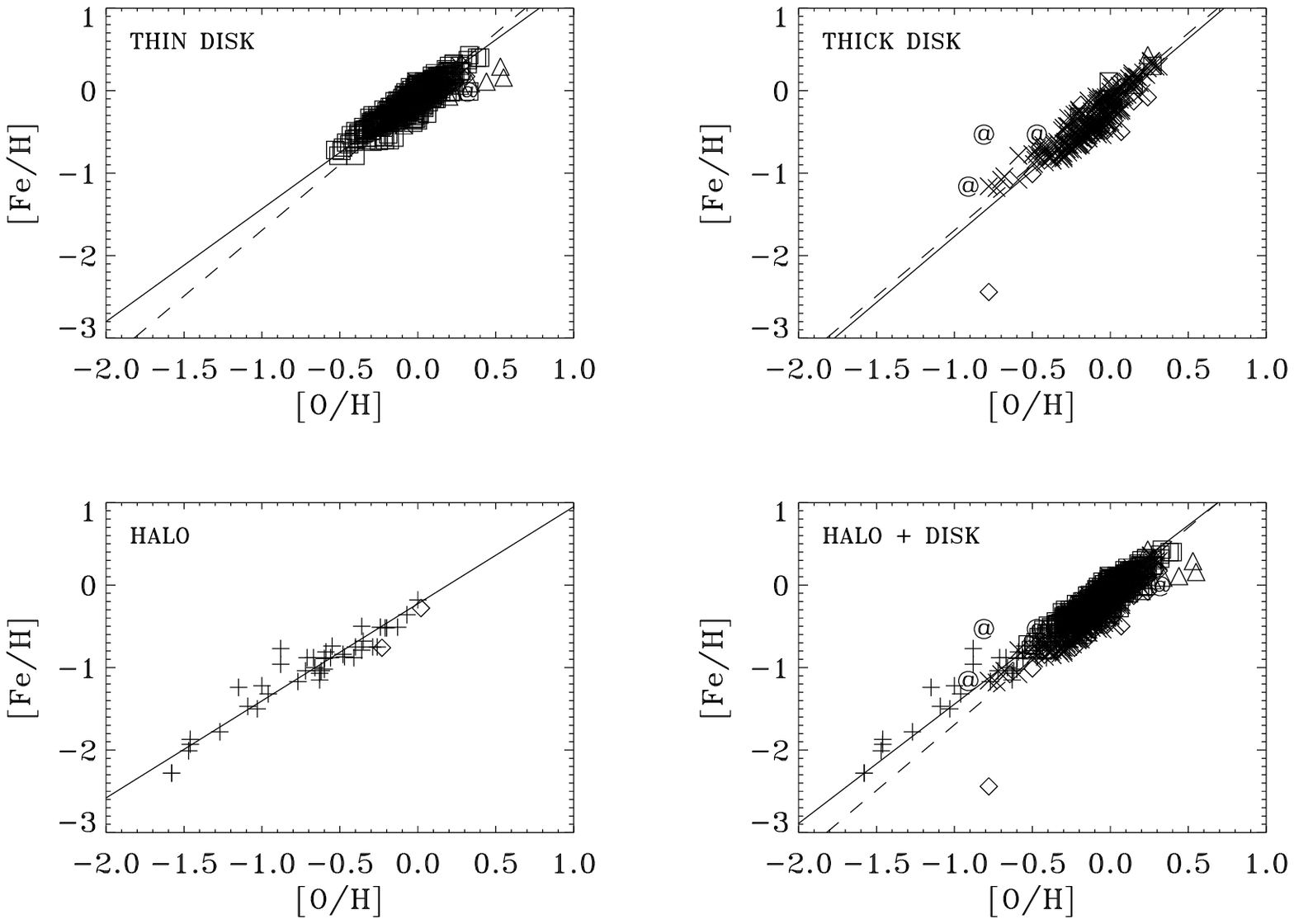}                      
\caption[ddbb]{Regression lines, [Fe/H$]=a[$O/H$]+b$, inferred for different
populations, ND (top left), KD (top right), HH (bottom left), HD (bottom
right).   Dashed lines relate to DD = ND + KD + KN.   Slope and intercept
values are taken from related
estimators listed in Table \ref{t:regre}.   The caption of the symbols is the
same as in Fig.\,\ref{f:ofehk}.}
\label{f:ofehkf}     
\end{center}       
\end{figure}

In summary, number abundances plotted in Fig.\,\ref{f:ofehk4} show a
linear trend as: [Fe/H] = $a$[O/H]$+b$  for HH, KD, ND, DD, HD populations.
The regression line has been determined for each
subsample using standard methods (e.g., Isobe et al. 1990; Caimmi 2011b,
2012b) and the results are listed in Table \ref{t:regre} and plotted in
Fig.\,\ref{f:ofehkf}.
\begin{table}
\caption{Regression line slope and intercept
estimators, $\hat{a}$ and $\hat{b}$, and
related dispersion estimators, $\hat{\sigma}_
{\hat{a}}$, and $\hat{\sigma}_{\hat{b}}$, for
different subsamples (sub)
.   The number $(N)$ of stars within each subsample is also listed.}
\label{t:regre}
\begin{center}
\begin{tabular}{lllllll} \hline
\multicolumn{1}{c}{$\hat{a}$}                &
\multicolumn{1}{c}{$\hat{\sigma}_{\hat{a}}$} &
\multicolumn{1}{c}{$-\hat{b}$}               &
\multicolumn{1}{c}{$\hat{\sigma}_{\hat{b}}$} &
\multicolumn{1}{l}{sub}                      &
\multicolumn{1}{c}{$N$}                      \\
\hline
           &              &              &              &    &     \\
1.1743D+00 & 4.7182D$-$02 & 2.2727D$-$01 & 3.6619D$-$02 & HH & 041 \\
1.6015D+00 & 4.4402D$-$02 & 1.6873D$-$01 & 9.0781D$-$03 & KD & 215 \\
1.3658D+00 & 2.9746D$-$02 & 7.0718D$-$02 & 5.1026D$-$03 & ND & 513 \\
1.5942D+00 & 4.8154D$-$02 & 9.5685D$-$02 & 7.8672D$-$03 & DD & 734 \\
1.4378D+00 & 5.0321D$-$02 & 1.0174D$-$01 & 7.9502D$-$03 & HD & 775 \\
\hline     
\end{tabular}
\end{center} 
\end{table}  
 From these results, it is evident that:
\begin{description}
\item[(1)\hspace{2.0mm}]
The regression line slope estimators, $\hat a$,  are not consistent%
\footnote{
The term ``consistent'' has to be intended in mathematical (instead of
statistical) sense, as a null intersection between intervals,
$\hat x\mp\hat\sigma_{\hat x}$, related to different populations, where $x$ is
a random variable.
}
 within
$\mp3\hat\sigma_{\hat a}$, for HH, KD, ND populations and the same holds for
DD, HD populations.
\item[(2)\hspace{2.0mm}]
The regression line slope estimators are not consistent with
the unit slope, within $\mp3\hat\sigma_{\hat a}$, regardless of the
population.
\item[(3)\hspace{2.0mm}]
The regression line intercept estimators, $\hat b$, 
are not consistent within $\mp3\hat\sigma_{\hat b}$ for HH, KD, ND populations
and the same holds for DD, HD populations.
\end{description}
These different slope and intercept values for different populations imply a
different chemical evolution for each region (HH, KD, ND, or DD if KD and
ND evolve as a single system). 

\section{Results} \label{resu}


The EDAD, $\psi_{\rm Q}=\log[\Delta N/(N\Delta\phi_{\rm Q})]$, inferred from
HH, KD, ND, DD subsamples, is listed in Tables \ref{t:OEDAD}-\ref{t:FeEDAD}
for oxygen and iron (Q = O, Fe), respectively.
Data are equally binned in [Q/H] taking $\Delta{\rm[Q/H]}=\Delta\log
\phi_{\rm Q}=1$dex, where the normalized abundance, $\phi_{\rm Q}$, is listed
in the first column, while the EDAD, $\psi_{\rm Q}$, and the number,
$\Delta N$, of subsample stars within the
abundance range, $\exp_{10}(\log\phi_{\rm Q}\mp\Delta\log\phi_{\rm Q})$, are
listed in the following columns for each related stellar population.
Uncertainties in $\psi_{\rm Q}$, $\Delta^\mp\psi_{\rm Q}$, are calculated as
Poissonian errors, which implies $\Delta^-\psi_{\rm Q}\to\infty$ for bins
populated by a single star, $\Delta N=1$.
\begin{table}
\caption{Oxygen empirical differential  abundance distribution (EDAD),
inferred from different subsamples, where $\delta=\Delta N$ to save 
space.   See text for further details.}
\label{t:OEDAD}
\begin{center}
\begin{tabular}{llrlrlrlr} \hline
\multicolumn{1}{c}{subsample:}
&\multicolumn{2}{c}{HH}
&\multicolumn{2}{c}{KD}
&\multicolumn{2}{c}{ND}
&\multicolumn{2}{c}{DD} \\
\multicolumn{1}{c}{$\phi_{\rm O}$} &
\multicolumn{1}{c}{$\phantom{0}\psi_{\rm O}$} &
\multicolumn{1}{c}{$\delta$} &
\multicolumn{1}{c}{$\phantom{0}\psi_{\rm O}$} &
\multicolumn{1}{c}{$\delta$} &
\multicolumn{1}{c}{$\phantom{0}\psi_{\rm O}$} &
\multicolumn{1}{c}{$\delta$} &
\multicolumn{1}{c}{$\phantom{0}\psi_{\rm O}$} &
\multicolumn{1}{c}{$\delta$} \\
\noalign{\smallskip}
\hline\noalign{\smallskip}
2.5286D$-$2 & $+$9.2507D$-$1 &  2 & $ $       $ $  &     & $ $       $ $  &     & $ $       $ $  &     \\
3.1833D$-$2 & $+$1.0012D$+$0 &  3 & $ $       $ $  &     & $ $       $ $  &     & $ $       $ $  &     \\
4.0075D$-$2 & $ $       $ $  &    & $ $       $ $  &     & $ $       $ $  &     & $ $       $ $  &     \\
5.0451D$-$2 & $+$3.2404D$-$1 &  1 & $ $       $ $  &     & $ $       $ $  &     & $ $       $ $  &     \\
6.3514D$-$2 & $+$2.2404D$-$1 &  1 & $ $       $ $  &     & $ $       $ $  &     & $ $       $ $  &     \\
7.9960D$-$2 & $+$1.2404D$-$1 &  1 & $ $       $ $  &     & $ $       $ $  &     & $ $       $ $  &     \\
1.0066D$-$1 & $+$5.0116D$-$1 &  3 & $ $       $ $  &     & $ $       $ $  &     & $ $       $ $  &     \\
1.2673D$-$1 & $+$2.2507D$-$1 &  2 & $ $       $ $  &     & $ $       $ $  &     & $ $       $ $  &     \\
1.5954D$-$1 & $-$1.7596D$-$1 &  1 & $-$8.9561D$-$1 &   1 & $ $       $ $  &     & $-$1.4289D$+$0 &   1 \\
2.0085D$-$1 & $+$3.2610D$-$1 &  4 & $-$5.1849D$-$1 &   3 & $ $       $ $  &     & $-$1.0518D$+$0 &   3 \\
2.5286D$-$1 & $+$5.2713D$-$1 &  8 & $-$7.9458D$-$1 &   2 & $ $       $ $  &     & $-$1.3278D$+$0 &   2 \\
3.1833D$-$1 & $-$1.7493D$-$1 &  2 & $-$4.1746D$-$1 &   6 & $-$8.7432D$-$1 &   5 & $-$6.8748D$-$1 &  11 \\
4.0075D$-$1 & $+$1.2301D$-$1 &  5 & $-$2.1643D$-$1 &  12 & $-$6.7329D$-$1 &  10 & $-$4.8645D$-$1 &  22 \\
5.0451D$-$1 & $+$3.7493D$-$1 &  2 & $+$3.5751D$-$2 &  27 & $-$1.2984D$-$1 &  44 & $-$7.7612D$-$2 &  71 \\
6.3514D$-$1 & $-$2.9884D$-$1 &  3 & $+$6.0689D$-$2 &  36 & $+$7.5216D$-$3 &  76 & $+$2.4208D$-$2 & 113 \\
7.9960D$-$1 & $-$5.7493D$-$1 &  2 & $+$9.4583D$-$2 &  49 & $-$9.5042D$-$3 &  92 & $+$2.3418D$-$2 & 142 \\
1.0066D$+$0 & $-$9.7596D$-$1 &  1 & $-$9.3553D$-$2 &  40 & $+$7.2836D$-$2 & 140 & $+$3.1201D$-$2 & 182 \\
1.2673D$+$0 & $ $       $ $  &    & $-$3.9767D$-$1 &  25 & $-$1.5626D$-$1 & 104 & $-$2.1828D$-$1 & 129 \\
1.5954D$+$0 & $ $       $ $  &    & $-$9.9252D$-$1 &   8 & $-$7.2922D$-$1 &  35 & $-$7.8542D$-$1 &  44 \\
2.0085D$+$0 & $ $       $ $  &    & $-$1.2175D$+$0 &   6 & $-$1.6743D$+$0 &   5 & $-$1.4497D$+$0 &  12 \\
2.5286D$+$0 & $ $       $ $  &    & $ $       $ $  &     & $-$2.1723D$+$0 &   2 & $-$2.3278D$+$0 &   2 \\
\hline
total:        &                  & 41 &                  & 215 &                  & 513 &                  & 734 \\
\noalign{\smallskip}      
\hline                                                       
\end{tabular}                                                
\end{center}                                                 
\end{table}                                                  
\begin{table}
\caption{Iron empirical differential  abundance distribution (EDAD), inferred 
from different subsamples, where $\delta=\Delta N$ to save 
space.   See text for further details.}
\label{t:FeEDAD}
\begin{center}
\begin{tabular}{llrlrlrlr} \hline
\multicolumn{1}{c}{subsample:}
&\multicolumn{2}{c}{HH}
&\multicolumn{2}{c}{KD}
&\multicolumn{2}{c}{ND}
&\multicolumn{2}{c}{DD} \\
\multicolumn{1}{c}{$\phi_{\rm Fe}$} &
\multicolumn{1}{c}{$\phantom{0}\psi_{\rm Fe}$} &
\multicolumn{1}{c}{$\delta$} &
\multicolumn{1}{c}{$\phantom{0}\psi_{\rm Fe}$} &
\multicolumn{1}{c}{$\delta$} &
\multicolumn{1}{c}{$\phantom{0}\psi_{\rm Fe}$} &
\multicolumn{1}{c}{$\delta$} &
\multicolumn{1}{c}{$\phantom{0}\psi_{\rm Fe}$} &
\multicolumn{1}{c}{$\delta$} \\
\noalign{\smallskip}
\hline\noalign{\smallskip}
5.0451D$-$3 & $+$1.6251D$+$0 &  2 & $ $       $ $  &     & $ $       $ $  &     & $ $       $ $  &     \\
6.3514D$-$3 & $ $       $ $  &    & $ $       $ $  &     & $ $       $ $  &     & $ $       $ $  &     \\
7.9960D$-$3 & $ $       $ $  &    & $ $       $ $  &     & $ $       $ $  &     & $ $       $ $  &     \\
1.0066D$-$2 & $+$1.0240D$+$0 &  1 & $ $       $ $  &     & $ $       $ $  &     & $ $       $ $  &     \\
1.2673D$-$2 & $+$1.2251D$+$0 &  2 & $ $       $ $  &     & $ $       $ $  &     & $ $       $ $  &     \\
1.5954D$-$2 & $+$8.2404D$-$1 &  1 & $ $       $ $  &     & $ $       $ $  &     & $ $       $ $  &     \\
2.0085D$-$2 & $ $       $ $  &    & $ $       $ $  &     & $ $       $ $  &     & $ $       $ $  &     \\
2.5286D$-$2 & $ $       $ $  &    & $ $       $ $  &     & $ $       $ $  &     & $ $       $ $  &     \\
3.1833D$-$2 & $+$8.2507D$-$1 &  2 & $ $       $ $  &     & $ $       $ $  &     & $ $       $ $  &     \\
4.0075D$-$2 & $ $       $ $  &    & $ $       $ $  &     & $ $       $ $  &     & $ $       $ $  &     \\
5.0451D$-$2 & $+$3.2404D$-$1 &  1 & $ $       $ $  &     & $ $       $ $  &     & $ $       $ $  &     \\
6.3514D$-$2 & $+$7.0116D$-$1 &  3 & $-$1.9458D$-$1 &   2 & $ $       $ $  &     & $-$7.2784D$-$1 &   2 \\
7.9960D$-$2 & $+$4.2507D$-$1 &  2 & $-$2.9458D$-$1 &   2 & $ $       $ $  &     & $-$8.2784D$-$1 &   2 \\
1.0066D$-$1 & $+$7.2301D$-$1 &  5 & $-$6.9561D$-$1 &   1 & $ $       $ $  &     & $-$1.2289D$+$0 &   1 \\
1.2673D$-$1 & $+$7.0219D$-$1 &  6 & $-$3.1849D$-$1 &   3 & $ $       $ $  &     & $-$8.5175D$-$1 &   3 \\
1.5954D$-$1 & $+$6.6914D$-$1 &  7 & $+$3.3484D$-$1 &  17 & $-$9.7226D$-$1 &   2 & $-$1.5012D$-$1 &  19 \\
2.0085D$-$1 & $+$2.5071D$-$2 &  2 & $+$2.5966D$-$1 &  18 & $-$7.7123D$-$1 &   4 & $-$1.8645D$-$1 &  22 \\
2.5286D$-$1 & $ $       $ $  &    & $+$1.0851D$-$1 &  16 & $-$3.5935D$-$1 &  13 & $-$1.6647D$-$1 &  29 \\
3.1833D$-$1 & $+$2.2301D$-$1 &  5 & $+$2.5154D$-$1 &  28 & $-$2.9454D$-$1 &  19 & $-$5.6773D$-$2 &  47 \\
4.0075D$-$1 & $-$5.7596D$-$1 &  1 & $+$1.1936D$-$1 &  26 & $+$1.2605D$-$1 &  63 & $+$1.2052D$-$1 &  89 \\
5.0451D$-$1 & $ $       $ $  &    & $-$1.5402D$-$2 &  24 & $+$5.2783D$-$2 &  67 & $+$3.0171D$-$2 &  91 \\
6.3514D$-$1 & $-$7.7596D$-$1 &  1 & $-$1.5319D$-$1 &  22 & $-$5.3748D$-$2 &  66 & $-$8.4388D$-$2 &  88 \\
7.9960D$-$1 & $ $       $ $  &    & $-$2.9458D$-$1 &  20 & $-$7.0202D$-$2 &  80 & $-$1.2887D$-$1 & 100 \\
1.0066D$+$0 & $ $       $ $  &    & $-$6.5422D$-$1 &  11 & $-$4.7986D$-$2 & 106 & $-$1.6068D$-$1 & 117 \\
1.2673D$+$0 & $ $       $ $  &    & $-$7.5422D$-$1 &  11 & $-$4.2510D$-$1 &  56 & $-$5.0280D$-$1 &  67 \\
1.5954D$+$0 & $ $       $ $  &    & $-$1.0505D$-$0 &   7 & $-$8.2613D$-$1 &  28 & $-$8.8480D$-$1 &  35 \\
2.0085D$+$0 & $ $       $ $  &    & $-$1.2966D$-$0 &   5 & $-$1.6743D$+$0 &   5 & $-$1.5289D$+$0 &  10 \\
2.5286D$+$0 & $ $       $ $  &    & $-$1.7946D$-$0 &   2 & $-$1.8712D$+$0 &   4 & $-$1.8507D$+$0 &   6 \\
\hline
total:        &                  & 41 &                  & 215 &                  & 513 &                  & 734 \\
\noalign{\smallskip}      
\hline                                                       
\end{tabular}                                                
\end{center}                                                 
\end{table}                                                  
\begin{figure}[t]  
\begin{center}      
\includegraphics[scale=0.8]{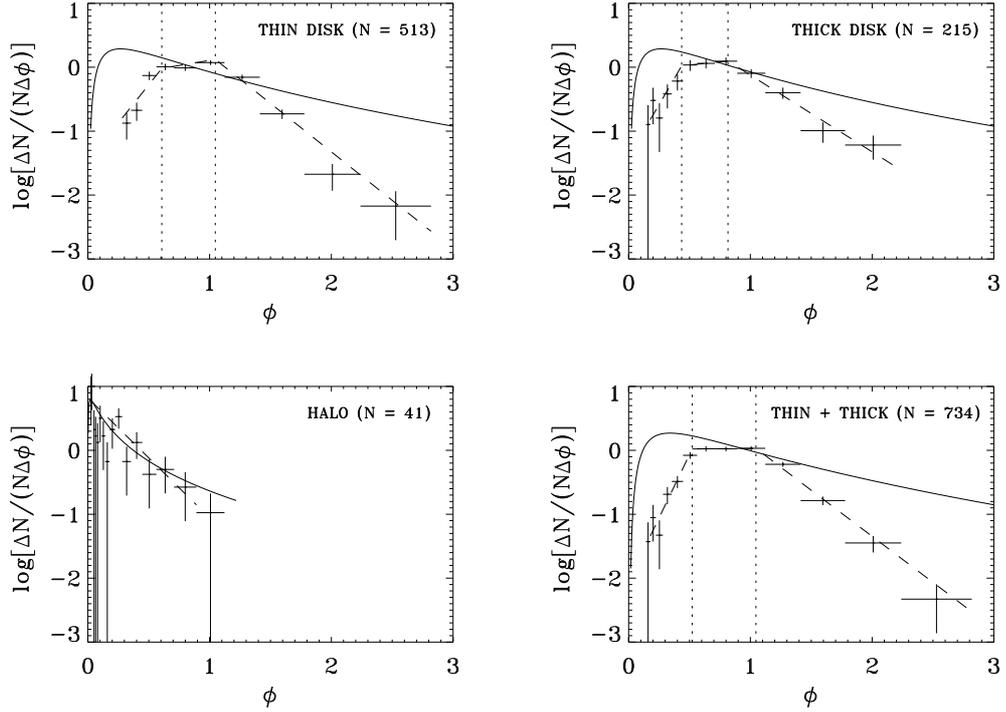}                      
\caption[ddbb]{Oxygen empirical differential abundance distribution (EDAD)
inferred from ND, KD, DD = ND + KD + KN, HH, subsamples (from top left
in clockwise sense).   Lower uncertainties
attaining the horizontal axis (decreasing up to negative infinity) relate to
bins populated by a single star.   Dashed straight lines represent regression
lines to points defining bins populated by at least two stars.   Transition
points between adjacent stages are marked as dotted vertical lines.
Full curves
represent oxygen theoretical differential abundance distribution (TDAD) due to 
intrinsic
scatter obeying a lognormal distribution with mean and variance inferred from
the data.
See text for further details.}
\label{f:eohk}     
\end{center}       
\end{figure}                                                                     
\begin{figure}[t]  
\begin{center}      
\includegraphics[scale=0.8]{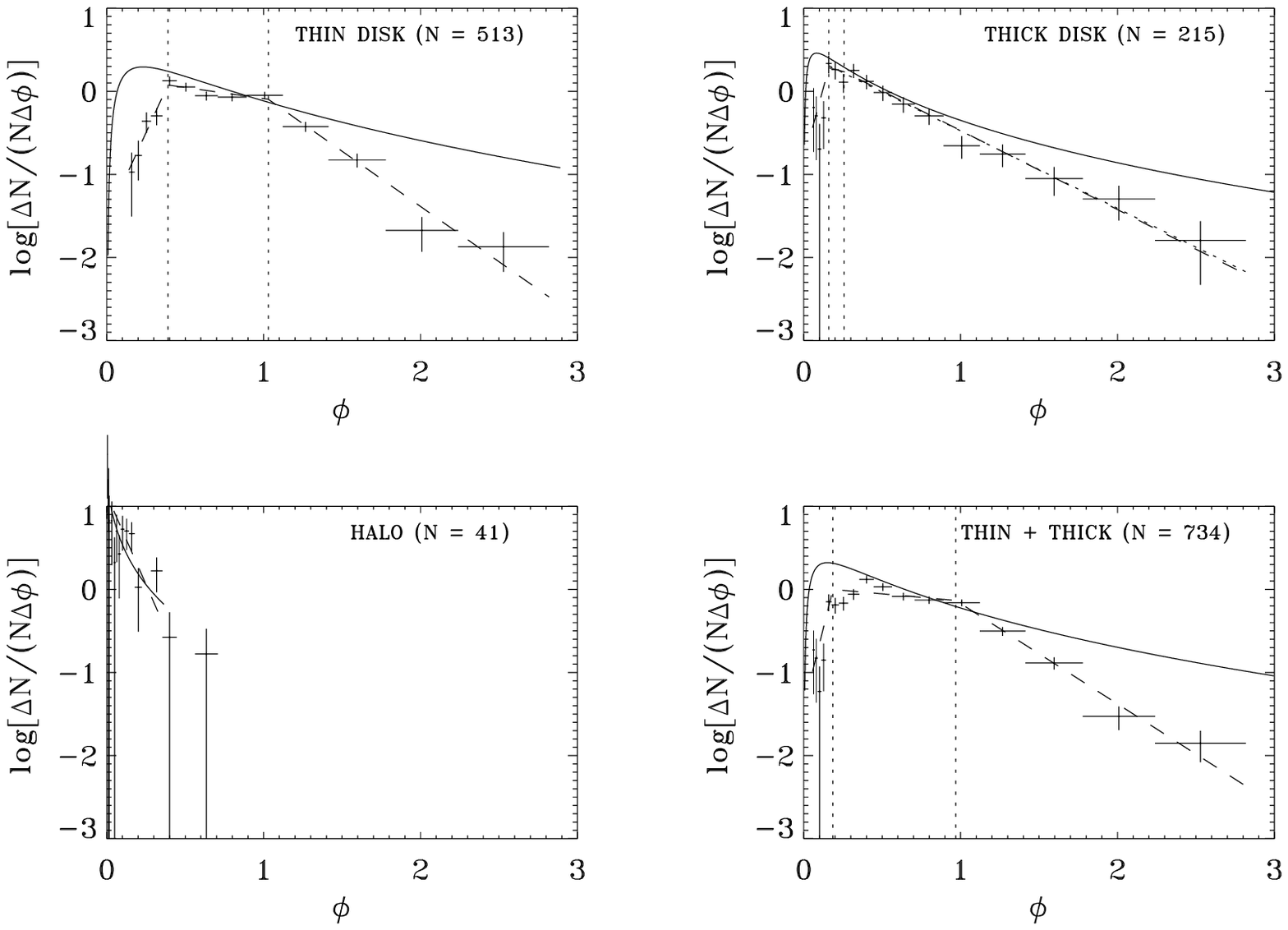}                      
\caption[ddbb]{Iron empirical differential abundance distribution (EDAD)
inferred from ND, KD, DD = ND + KD + KN, HH  subsamples (from top left
in clockwise sense).   Lower uncertainties
attaining the horizontal axis (decreasing up to negative infinity) relate to
bins populated by a single star.   Dashed straight lines represent regression
lines to points defining bins populated by at least two stars.   The dotted
straight line on top right panel represents the regression line for the whole
declining part of the EDAD.   Transition points between adjacent stages are
marked as dotted vertical lines.
Full curves represent iron theoretical
differential abundance distribution (TDAD) due to intrinsic
scatter obeying a lognormal distribution with mean and variance inferred from
the data.
See text for further details.}
\label{f:efehk}     
\end{center}       
\end{figure}                                                                     

The above mentioned EDAD is plotted in Figs.\,\ref{f:eohk}-\ref{f:efehk} for O
and Fe, respectively, separating results for each stellar population (ND, KD,
DD = ND + KD + KN, and HH) in a different panel.   Lower uncertainties
attaining the horizontal axis (decreasing down to negative infinity) relate to
bins populated by a single star.   Regression lines (dashed lines) have been
performed to points in each stage (defining bins populated by at least two
stars).

Arithmetic mean and rms error can be inferred from the EDAD as (C13a; C13b):
\begin{lefteqnarray}
\label{eq:lgfi}
&& \overline{\log\phi}=\overline{\rm[Q/H]}=\frac1N\sum_{i=1}^N
{\rm[Q/H]}_i~~; \\
\label{eq:slfi}
&& \sigma_{\log\phi}=\sigma_{\rm[Q/H]}=\left\{\frac1{N-1}\sum_{i=1}^N
\left({\rm[Q/H]}_i-\overline{\rm[Q/H]}\right)^2\right\}^{1/2}~~;
\end{lefteqnarray}
where Q = O, Fe, and $N$ is the subsample population.   The results are listed
in Table \ref{t:parga} for subsamples, HH, KD, ND, DD.
\begin{table*}
\caption{Star number, $N$, mean abundance, $\overline{[{\rm Q/H}]}$, rms error,
$\sigma_{[{\rm Q/H}]}$, Q = O, Fe,
inferred for different subsamples (sub).
%
%
See text for further details.}
\label{t:parga}
\begin{center}
\begin{tabular}{rlllll} \hline
\multicolumn{1}{c}{$N$} &
\multicolumn{1}{c}{$\phantom{-}\overline{[{\rm O/H}]}$} &
\multicolumn{1}{c}{$\sigma_{[{\rm O/H}]}$} &
\multicolumn{1}{c}{$\phantom{-}\overline{[{\rm Fe/H}]}$} &
\multicolumn{1}{c}{$\sigma_{[{\rm Fe/H}]}$} &
\multicolumn{1}{c}{sub} \\
\hline
 41 & $-$0.6915 & 0.4274 & $-$1.0393 & 0.5020 & HH \\
215 & $-$0.1172 & 0.1979 & $-$0.3564 & 0.3174 & KD \\
513 & $-$0.0511 & 0.1580 & $-$0.1401 & 0.2161 & ND \\
734 & $-$0.0699 & 0.1733 & $-$0.2028 & 0.2690 & DD \\
\hline
\end{tabular}
\end{center}                      
\end{table*}                       

\section{Discussion} \label{disc}

The following main points shall be discussed throughout the current section,
concerning (1) abundance distributions and their interpretation within the
framework of simple MCBR models, where special effort is devoted to inflow and
outflow considerations, and evolution of different regions; (2) iron yields
related to both sp and np processes, and oxygen-to-iron yield ratios.

\subsection{Abundance distributions and their interpretation}
\subsubsection{Oxygen and iron TDAD} \label{OFTDAD}

The theoretical differential abundance distribution (TDAD),\linebreak
$\psi_{\rm Q}=\log[\diff N/(N\diff\phi_{\rm Q})]$, predicted by simple MCBR
chemical evolution
models, is a  broken line (e.g., Caimmi 2011a, 2012a). The straight
line defined by each segment can be expressed as (Pagel 1989):
\begin{lefteqnarray}
\label{eq:psit}
&& \psi_{\rm Q}=
\alpha_{\rm Q}\phi_{\rm Q}+\beta_{\rm Q}~~; \\
\label{eq:fQfH}
&& \phi_{\rm Q}=\frac{Z_{\rm Q}}{(Z_{\rm Q})_\odot}~~;\qquad\phi_{\rm H}=\frac
{X}{X_\odot}~~;
\end{lefteqnarray}
where $X=Z_{\rm H}$ according to the standard notation, Q is a selected
element heavier than He (O and Fe in our case), $\log\phi_{\rm Q}={\rm [Q/H]}$
as a good approximation and $(Z_{\rm Q})_\odot$ is the solar abundance.

The explicit expression of TDAD slope and intercept reads (e.g., Caimmi 2011a,
2012a):
\begin{lefteqnarray}
\label{eq:salpha}
&& \alpha_{\rm Q}=-\frac1{\ln10}\frac{(Z_{\rm Q})_\odot}{\hat p_{\rm Q}}
(1+\kappa)~~; \\
\label{eq:sbeta}
&& \beta_{\rm Q}=\log\left[\frac{\mu_{\rm i}}{\mu_{\rm i}-\mu_{\rm f}}(-\ln10)
\alpha_{\rm Q}\right]-\alpha_{\rm Q}(\phi_{\rm Q})_{\rm i}~~;
\end{lefteqnarray}
where $\hat p_{\rm Q}$ is the
yield per stellar generation (e.g., Pagel and Patchett 1975) of
the sp element, Q, and $\kappa$ is the flow parameter,
$\mu$ the active (i.e. available for star formation) gas mass 
fraction, i and f denote values at the start (minimum Q abundance) and the end
(maximum Q abundance) of the stage considered, respectively.

More specifically, the following flow regimes can be defined (e.g., Caimmi
2011a, 2012a): outflow $(\kappa>0)$, where star formation is lowered
(Hartwick 1976); stagnation $(\kappa=0)$, where star formation is neither
lowered nor enhanced; weak inflow $(-1<\kappa<0)$, where star formation
is weakly enhanced and gas mass fraction
monotonically decreases in time (Caimmi 2007); steady inflow $(\kappa=-1)$,
where star formation is moderately enhanced and gas mass fraction remains
unchanged; strong inflow $(\kappa<-1)$, where star formation is strongly
enhanced and gas mass fraction monotonically increases in time.

Owing to Eq.\,(\ref{eq:salpha}), strong inflow, steady inflow, weak inflow
or stagnation or outflow, imply positive, null, negative TDAD slope,
respectively.  In current $\Lambda$CDM scenarios, strong inflow may safely be
related to an assembling stage of galaxy evolution, steady state inflow to a
formation stage, weak inflow or stagnation or outflow to a next evolution
stage (e.g., Finlator and Dav\'e 2008; Dav\'e et al. 2011a,b, 2012).

\subsubsection{Oxygen and iron EDAD} \label{OFEDAD}

Within the framework of simple MCBR chemical evolution models, the TDAD of sp
elements may be considered in connection with different stages of evolution.
More specifically, the early stage with positive slope relates to
strong gas inflow (SI), where the gas mass fraction is increasing in time; the
middle stage with nearly flat slope relates to (nearly) steady state gas
inflow (SS),
where the gas mass fraction remains (more or less) unchanged in time; the late
stage with negative slope relates to weak gas inflow or outflow (WI), where
the gas mass fraction is decreasing in time.

Oxygen and iron EDAD shown in Fig.\,\ref{f:eohk} and \ref{f:efehk},
respectively, for disk populations is characterized
by three distinct stages exhibiting different linear trends.
For each stage, regression lines shown in Figs.\,\ref{f:eohk}-\ref{f:efehk}
have been determined using standard methods (e.g., Isobe et al. 1990; Caimmi
2011b, 2012b), leaving aside points related to bins containing a single star,
where $\Delta^-\psi_{\rm Q}\to\infty$.   The regression procedure has been
performed on HH, KD, ND, DD subsamples and the results are shown in Tables
\ref{t:rego}-\ref{t:regfe} for oxygen and iron, respectively.   The
transition points between adjacent stages are determined as intersections of
related regression lines and the results are shown in Table \ref{t:pint} for
both oxygen and iron, which are marked as dotted vertical lines in
Figs.\,\ref{f:eohk} and \ref{f:efehk}, respectively.
\begin{table}
\caption{Regression line slope and intercept
estimators, $\hat{\alpha}_{\rm O}$ and $\hat{\beta}_{\rm O}$, and
related dispersion estimators, $\hat{\sigma}_
{\hat{\alpha}_{\rm O}}$, and $\hat{\sigma}_{\hat{\beta}_{\rm O}}$,
for regression models applied to the oxygen empirical differential
abundance distribution (EDAD) plotted in Fig.\,\ref{f:eohk},
related to different regions (reg).
The method has dealt with
each stage (S) separately: A - assembling, F - formation, E - evolution.
A single stage (E) has been
considered for halo population due to the incompleteness of related subsample.
Bins containing a single star are not considered in the regression.}
\label{t:rego}
\begin{center}
\begin{tabular}{llllll} \hline
\multicolumn{1}{l}{S} &
\multicolumn{1}{c}{$\hat{\alpha}_{\rm O}$} &
\multicolumn{1}{c}{$\hat{\sigma}_{\hat{\alpha}_{\rm O}}$} &
\multicolumn{1}{c}{$\hat{\beta}_{\rm O}$} &
\multicolumn{1}{c}{$\hat{\sigma}_{\hat{\beta}_{\rm O}}$} &
\multicolumn{1}{l}{reg} \\
\hline
  &                 &              &                 &              &    \\
A & $+$3.2643 E$-$0 & 1.8045 E$-$0 & $-$1.4001 E$-$0 & 5.0690 E$-$1 & KD \\
A & $+$2.3871 E$-$0 & 7.5607 E$-$1 & $-$1.4681 E$-$0 & 4.2787 E$-$1 & ND \\
A & $+$3.9590 E$-$0 & 1.0948 E$-$0 & $-$2.0445 E$-$0 & 3.3269 E$-$1 & DD \\
F & $+$1.9851 E$-$1 & 4.4600 E$-$3 & $-$6.4657 E$-$2 & 2.7360 E$-$3 & KD \\
F & $+$2.0642 E$-$1 & 1.3444 E$-$1 & $-$1.4400 E$-$1 & 1.1628 E$-$1 & ND \\
F & $+$1.6397 E$-$2 & 1.2453 E$-$2 & $+$1.2886 E$-$2 & 9.6177 E$-$3 & DD \\
E & $-$1.2001 E$-$0 & 1.0914 E$-$1 & $+$1.0764 E$-$0 & 1.2480 E$-$1 & KD \\
E & $-$1.4891 E$-$0 & 1.8978 E$-$1 & $+$1.6313 E$-$0 & 2.3646 E$-$1 & ND \\
E & $-$1.5392 E$-$0 & 8.5043 E$-$2 & $+$1.6397 E$-$0 & 1.2252 E$-$1 & DD \\
E & $-$1.8764 E$-$0 & 3.0416 E$-$1 & $+$8.1914 E$-$1 & 1.1220 E$-$1 & HH \\
\hline       
\end{tabular}
\end{center} 
\end{table}  
\begin{table}
\caption{Regression line slope and intercept
estimators, $\hat{\alpha}_{\rm Fe}$ and $\hat{\beta}_{\rm Fe}$, and
related dispersion estimators, $\hat{\sigma}_
{\hat{\alpha}_{\rm Fe}}$, and $\hat{\sigma}_{\hat{\beta}_{\rm Fe}}$,
for regression models applied to the iron empirical differential
abundance distribution (EDAD) plotted in Fig.\,\ref{f:efehk},
related to different regions (reg).
The method has dealt with
each stage (S) separately: A - assembling, F - formation, E - evolution.
A single stage (E) has been
considered for halo population due to the incompleteness of related subsample.
For KD subsample, stages F and E are also considered as a single stage, FE.
Bins containing a single star are not considered in the regression.}
\label{t:regfe}
\begin{center}
\begin{tabular}{llllll} \hline
\multicolumn{1}{l}{S} &
\multicolumn{1}{c}{$\hat{\alpha}_{\rm Fe}$} &
\multicolumn{1}{c}{$\hat{\sigma}_{\hat{\alpha}_{\rm Fe}}$} &
\multicolumn{1}{c}{$\hat{\beta}_{\rm Fe}$} &
\multicolumn{1}{c}{$\hat{\sigma}_{\hat{\beta}_{\rm Fe}}$} &
\multicolumn{1}{l}{reg} \\
\hline
   &                 &              &                 &              &    \\
A  & $+$6.9323 E$-$0 & 3.2108 E$-$0 & $-$8.2705 E$-$1 & 4.5176 E$-$1 & KD \\
A  & $+$4.1103 E$-$0 & 5.0477 E$-$1 & $-$1.5270 E$-$0 & 1.7900 E$-$1 & ND \\
A  & $+$7.5777 E$-$0 & 3.3920 E$-$0 & $-$1.4135 E$-$0 & 4.8156 E$-$1 & DD \\
F  & $-$4.8416 E$-$1 & 7.8092 E$-$1 & $+$3.5966 E$-$1 & 1.9595 E$-$1 & KD \\
F  & $-$2.5801 E$-$1 & 1.1309 E$-$1 & $+$1.7609 E$-$1 & 8.4070 E$-$2 & ND \\
F  & $-$1.6209 E$-$1 & 1.2974 E$-$1 & $+$2.9866 E$-$2 & 8.5168 E$-$2 & DD \\
E  & $-$9.5208 E$-$1 & 4.8752 E$-$2 & $+$4.8016 E$-$1 & 4.3781 E$-$2 & KD \\
E  & $-$1.3329 E$-$0 & 1.0979 E$-$1 & $+$1.2826 E$-$0 & 1.4090 E$-$1 & ND \\
E  & $-$1.2115 E$-$0 & 7.2617 E$-$2 & $+$1.0469 E$-$0 & 9.5796 E$-$2 & DD \\
E  & $-$3.1707 E$-$0 & 8.8934 E$-$1 & $+$1.0858 E$-$0 & 1.4100 E$-$1 & HH \\
FE & $-$9.3301 E$-$1 & 4.2058 E$-$2 & $+$4.5619 E$-$1 & 3.3064 E$-$2 & KD \\
\hline
\end{tabular}
\end{center} 
\end{table}  
\begin{table}
\caption{Transition (trans) points between adjacent
stages, as determined from the intersection of related regression lines, 
$(\phi_{\rm Q},\psi_{\rm Q})$, Q = O, Fe, for the empirical differential 
abundance distribution (EDAD) plotted in Figs.\,\ref{f:eohk}-\ref{f:efehk},
respectively, with regard to HH, KD, ND, DD
subsample (sub).   For KD subsample, stages F and E are also considered as a
single stage, FE, with regard to iron.  Regression lines outside the domain of
subsample abundance, denoted as stage O, are assumed to be vertical lines.}
\label{t:pint}
\begin{center}
\begin{tabular}{cllllll} \hline
\multicolumn{1}{l}{trans} &
\multicolumn{1}{c}{$\phi_{\rm O}$} &
\multicolumn{1}{c}{$\psi_{\rm O}$} &
\multicolumn{1}{c}{$\phi_{\rm Fe}$} &
\multicolumn{1}{c}{$\psi_{\rm Fe}$} &
\multicolumn{1}{c}{sub}  \\
\hline
     &              &                 &             &                &    \\
O-E  & 2.2387 E$-$2 & $+$7.7714 E$-$1 & 4.4668E$-$2 & $+$9.4414E$-$1 & HH \\ 
E-O  & 8.9125 E$-$1 & $-$8.5318 E$-$1 & 3.5481E$-$1 & $-$3.9241E$-$1 & HH \\ 
     &              &                 &             &                &    \\
O-A  & 1.7783 E$-$1 & $-$8.1960 E$-$1 & 5.6234E$-$2 & $-$4.3722E$-$1 & KD \\ 
A-F  & 4.3559 E$-$1 & $+$2.1810 E$-$2 & 1.6001E$-$1 & $+$2.8219E$-$1 & KD \\
F-E  & 8.1590 E$-$1 & $+$9.7304 E$-$2 & 2.5752E$-$1 & $+$2.3498E$-$1 & KD \\ 
E-O  & 2.2387 E$-$0 & $-$1.6102 E$-$0 & 2.8184E$-$0 & $-$2.2032E$-$0 & KD \\ 
A-FE &              &                 & 1.6315E$-$1 & $+$3.0397E$-$1 & KD \\
     &              &                 &             &                &    \\
O-A  & 2.8184 E$-$1 & $-$7.9533 E$-$1 & 1.4125E$-$1 & $-$9.4641E$-$1 & ND \\ 
A-F  & 6.0720 E$-$1 & $-$1.8662 E$-$2 & 3.8987E$-$1 & $+$7.5496E$-$2 & ND \\
F-E  & 1.0470 E$-$0 & $+$1.2229 E$-$1 & 1.0294E$-$0 & $-$9.6202E$-$2 & ND \\ 
E-O  & 2.8184 E$-$0 & $-$2.5657 E$-$0 & 2.8184E$-$0 & $-$2.4742E$-$0 & ND \\ 
     &              &                 &             &                &    \\
O-A  & 1.7783 E$-$1 & $-$1.3404 E$-$0 & 5.6234E$-$2 & $-$9.8735E$-$1 & DD \\ 
A-F  & 5.2182 E$-$1 & $+$2.1442 E$-$2 & 1.8648E$-$1 & $-$3.6075E$-$4 & DD \\
F-E  & 1.0458 E$-$0 & $+$3.0034 E$-$2 & 9.6920E$-$1 & $-$1.3351E$-$1 & DD \\ 
E-O  & 2.8184 E$-$0 & $-$2.5252 E$-$0 & 2.8184E$-$0 & $-$2.3674E$-$0 & DD \\ 
\hline                                                         
\end{tabular}
\end{center} 
\end{table}  

By comparing with the theoretical MCBR model expectations, these trends may
be interpreted as follows:  the early stage with
positive slope may safely be related to assembling
(A), the middle stage with nearly flat slope to formation (F), the
late stage with negative slope to evolution (E).   The last stage
could be in connection with different trends implying two fitting
straight lines which, in general, form a knee.

In this view, A stage is characterized by SI inflow regime,
F stage by SS inflow regime, E stage by WI inflow
(to be intended as including outflow)
regime, as shown in Fig.\,\ref{f:eohk} for oxygen EDAD.   Iron
EDAD exhibits a similar trend with respect to oxygen and, in addition, a peak
related to F stage, as shown in Fig.\,\ref{f:efehk}.   If delayed recycling
via SNIa progenitors is the main
process which makes iron np element, then the above
mentioned peak can be related to the onset of SNIa events.

In summary, the following informations can be inferred from oxygen and iron
EDAD via Table \ref{t:pint}: (i) the transition from SI to SS inflow regime
took place at [O/H$]\approx-0.36$ for the thick disk and [O/H$]\approx-0.22$
for the thin disk; the transition from SS to WI inflow regime took place at
[O/H$]\approx-0.09$ for the thick disk and [O/H$]\approx+0.02$ for the thin
disk; (ii) the onset of SNIa events took place at [Fe/H$]\approx-0.80$ for
the thick disk and [Fe/H$]\approx-0.40$ for the thin disk.   No conclusion can
be drawn for the halo, due to the incompleteness of related subsample.

With regard to KD subsample, oxygen EDAD shows a similar trend to the one
found in an earlier investigation (Caimmi 2012a) for a sample of 133 thick
disk stars from Ram{\'\i}rez et al. (2007).   However, when we compare the
present results with those from C13b, obtained with the sample of Ram{\'\i}rez
et al. (2012), we see that, while iron EDAD exhibits a similar trend to that
one, within the common abundance range, the contrary holds for oxygen EDAD.
More specifically, fitting straight lines
within the common abundance range ([O/H$]<-0.05$) show a negative slope
instead
of positive or close to zero as outlined in Fig.\,\ref{f:eohk}, top right
panel $(\phi_{\rm O}<0.9)$.   The above mentioned discrepancy is probably
owing to the
combined effect of poor subsample (16 stars compared with the 133 from
Ram{\'\i}rez et al. 2007), biases against high iron abundance, larger
oxygen abundance scatter with respect to iron abundance.

\subsubsection{Nonlinear iron TDAD} \label{FeTDAD}

Let Q be a selected element for which both the empirical [Q/H]-[O/H] relation:
\begin{equation}
\label{eq:QO}
[{\rm Q/H}]=a_{\rm Q}[{\rm O/H}]+b_{\rm Q}~~;
\end{equation}
and oxygen EDAD have been inferred from a selected star sample.
Using the relation between number abundance and mass abundance (e.g., Caimmi
2007; C13b):
\begin{lefteqnarray}
\label{eq:fiQH}
&& \log\frac{\phi_{\rm Q}}{\phi_{\rm H}}={\rm[Q/H]}~~;
\end{lefteqnarray}
after little algebra  Eq.\,(\ref{eq:QO}) translates into:
\begin{lefteqnarray}
\label{eq:QOpl}
&& \phi_{\rm Q}=B_{\rm Q}\phi_{\rm H}^{1-a_{\rm Q}}\phi_{\rm O}^{a_{\rm Q}}~~;
\qquad B_{\rm Q}=\exp_{10}(b_{\rm Q})~~;
\end{lefteqnarray}
where $\phi_{\rm H}$ is the hydrogen abundance normalized to the solar value,
which is expected to change only slightly in time and may safely be assumed as
constant.
Accordingly, a differentiation on both sides of Eq.\,(\ref{eq:QOpl}) yields:
\begin{lefteqnarray}
\label{eq:dQO}
&& \diff\phi_{\rm Q}=B_{\rm Q}a_{\rm Q}\left(\frac
{\phi_{\rm O}}{\phi_{\rm H}}\right)^{a_{\rm Q}-1}\diff\phi_{\rm O}~~;
\end{lefteqnarray}
where $\diff\phi_{\rm Q}$ and $\diff\phi_{\rm O}$ may be conceived as
infinitely thin bins centered on $\phi_{\rm Q}$ and $\phi_{\rm O}$,
respectively, both containing an equal infinitesimal number of stars,
$\diff N$.

Finally, the TDAD of the element, Q, may be expressed as:
\begin{lefteqnarray}
\label{eq:psQ1}
&& \psi_{\rm Q}=\log\left(\frac{\diff N}{N\diff\phi_{\rm Q}}\right)=
\log\left(\frac{\diff N}{N\diff\phi_{\rm O}}
\frac{\diff\phi_{\rm O}}{\diff\phi_{\rm Q}}\right)=
\psi_{\rm O}+\log\left(\frac{\diff\phi_{\rm O}}{\diff\phi_{\rm Q}}\right)~~;
\end{lefteqnarray}
and the substitution of Eqs.\,(\ref{eq:psit}) and (\ref{eq:dQO}) into
(\ref{eq:psQ1}) after some algebra yields:
\begin{lefteqnarray}
\label{eq:psQ2}
&& \psi_{\rm Q}=\alpha_{\rm O}\phi_{\rm O}+\beta_{\rm O}-b_{\rm Q}-
\log a_{\rm Q}-(a_{\rm Q}-1)\log\left(\frac{\phi_{\rm O}}{\phi_{\rm H}}
\right)~~;
\end{lefteqnarray}
which, in terms of Q abundance, by use of Eq.\,(\ref{eq:QOpl}), after some
algebra translates into:
\begin{lefteqnarray}
\label{eq:psQ3}
&& \psi_{\rm Q}=\alpha_{\rm Q}\left(\frac{\phi_{\rm Q}}{\phi_{\rm H}}\right)^
{1/a_{\rm Q}}\phi_{\rm H}+\beta_{\rm Q}-(a_{\rm Q}-1)
\log\left(\frac{\phi_{\rm Q}}{\phi_{\rm H}}\right)^{1/a_{\rm Q}}~~; \\
\label{eq:ppQ}
&& \alpha_{\rm Q}=\frac{\alpha_{\rm O}}{B_{\rm Q}^{1/a_{\rm Q}}}~~;\qquad
\beta_{\rm Q}=\beta_{\rm O}-\log a_{\rm Q}-\frac{b_{\rm Q}}{a_{\rm Q}}~~;
\end{lefteqnarray}
where hydrogen abundance, $\phi_{\rm H}$, may safely be put equal to unity.
If the element, Q, is sp, $a_{\rm Q}=1$, then Eq.\,(\ref{eq:psQ3})
reduces to (\ref{eq:psit}).

\begin{figure}[t]  
\begin{center}      
\includegraphics[scale=0.8]{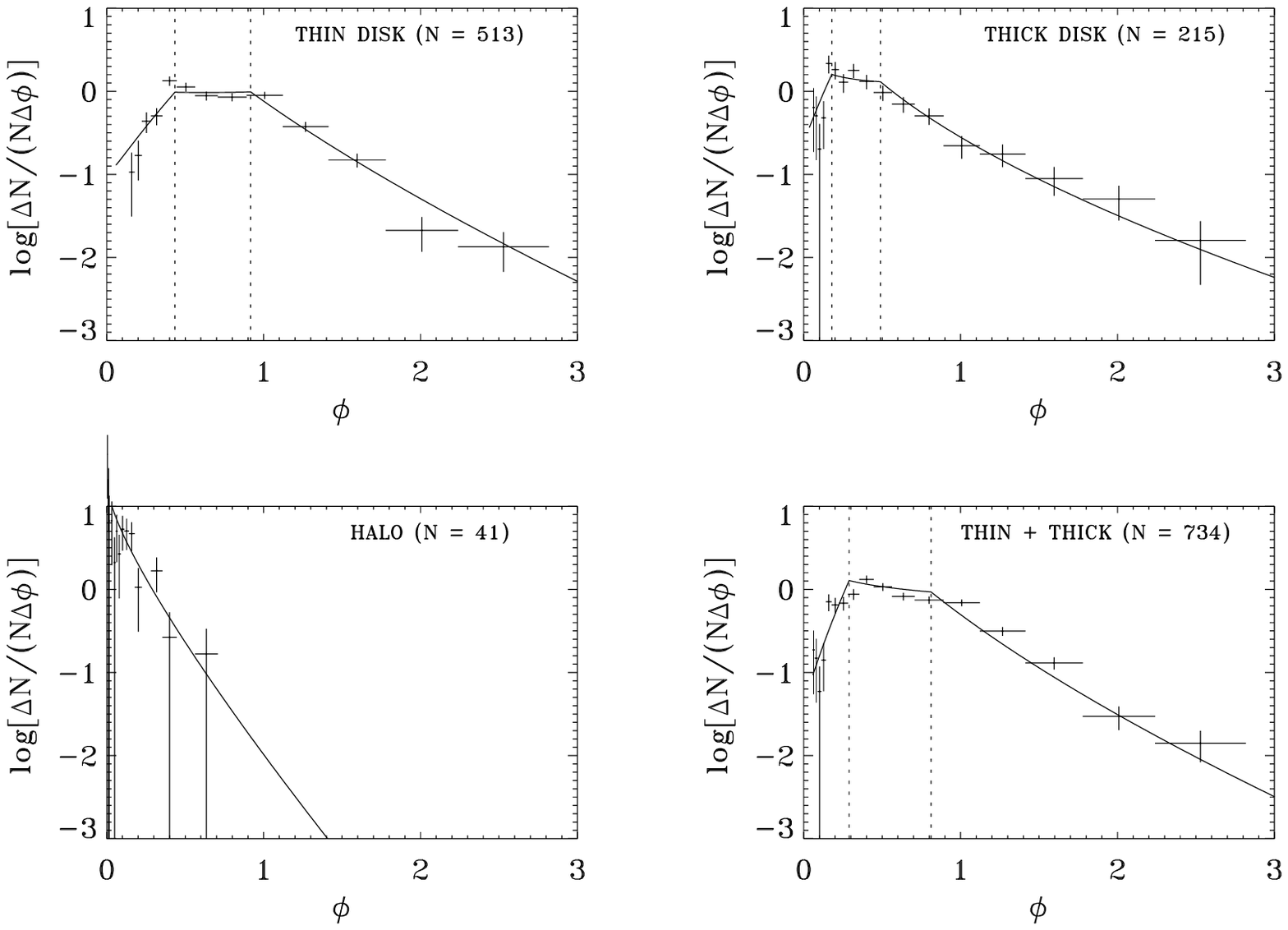}                      
\caption[ddbb]{Iron theoretical differential abundance distribution (TDAD)
inferred from the empirical [Fe/H]-[O/H] relation and oxygen empirical
differential abundance distribution (EDAD), with regard to ND, KD, DD, HH
subsamples (from top left in clockwise sense).   Iron EDAD is also shown for
comparison (captions as in Fig.\,\ref{f:efehk}).   Dotted vertical lines
mark steady state (SS) inflow regime inferred from oxygen EDAD.   See text for
further details.}
\label{f:efeo4}     
\end{center}       
\end{figure}

The special case of iron, related to HH, KD, ND, DD subsamples, is plotted
in Fig.\,\ref{f:efeo4} using the results listed in Tables \ref{t:regre} and
\ref{t:rego}, and compared with its empirical counterpart shown in
Fig.\,\ref{f:efehk}.
The SS inflow regime, inferred from oxygen EDAD via Eq.\,(\ref{eq:QOpl}), is
marked by dotted vertical lines.   An inspection of Fig.\,\ref{f:efeo4} shows
iron TDAD, expressed by Eq.\,(\ref{eq:psQ3}) where Q = Fe, $\phi_{\rm H}=1$,
and the values of coefficients are inferred from Tables \ref{t:regre} and
\ref{t:rego} via
Eq.\,(\ref{eq:ppQ}), can be approximated, to a first extent, by regression
lines to iron EDAD for different inflow regimes, shown in Fig.\,\ref{f:efehk}.
Therefore simple MCBR chemical evolution models may safely be used for the
description of iron
chemical evolution, with regard to subsamples under consideration.

\subsubsection{Inflow/outflow rate for different populations}
\label{fior}

With regard to oxygen, which may safely be conceived as sp element within
the framework of simple MCBR chemical evolution models, the assumption of
universal stellar IMF
implies constant oxygen yield, $\hat p_{\rm O}$, for different populations.
Though products of stellar nucleosynthesis  vary with the initial metallicity
(e.g., Woosley and Weaver 1995), still oxygen and iron yields per star
generation show a mild
dependence (e.g., Yates et al. 2013) which can be neglected to a first extent.

Accordingly, the following relation (C13b)  is easily inferred from
Eq.\,(\ref{eq:salpha}):
\begin{lefteqnarray}
\label{eq:ap12}
&& \frac{(\alpha_{\rm O})_{\rm XY}}{(\alpha_{\rm O})_{\rm WZ}}=\frac
{1+\kappa_{\rm XY}}{1+\kappa_{\rm WZ}}~~;
\end{lefteqnarray}
where $\kappa$ is the flow parameter, proportional to inflow ($\kappa<0$) or
outflow ($\kappa>0$) rate, as explained, and XY, WZ, identify subsamples
under consideration.
For constant effective yield, $\hat p_{\rm O}/(1+\kappa)$, i.e. constant flow
parameter,
$\kappa$, the $\alpha_{\rm O}$ ratio related to different regions remains
unchanged.   Then $\alpha_{\rm O}$ ratios different from unity imply different
inflow or outflow rate for different regions.

Oxygen TDAD slopes, $\alpha_{\rm O}$, are inferred from the data, which
implies constraints on flow parameters related to different environments, via
Eq.\,(\ref{eq:ap12}).
Keeping in mind regression lines
exhibit slope of equal sign for a selected stage, as shown in Table
\ref{t:rego}, Eq.\,(\ref{eq:ap12}) discloses that related flow parameters
must be both larger/lower than or equal to $-1$, which implies the additional
condition, $(\alpha_{\rm O})_{\rm XY}/(\alpha_{\rm O})_{\rm WZ}>0$, holds
for a selected stage according to the results listed in Table \ref{t:rego}.
Little algebra shows the correlation between slope ratio and flow
parameter ratio, for assigned values of $\kappa_{\rm WZ}$, as:
\begin{leftsubeqnarray}
\slabel{eq:ap13a}
&& \frac{(\alpha_{\rm O})_{\rm XY}}{(\alpha_{\rm O})_{\rm WZ}}\leqgr1~~;
\qquad
\frac{\kappa_{\rm XY}}{\kappa_{\rm WZ}}\leqgr1~~;\quad\kappa_{\rm WZ}\leq-1
~~;
\quad\kappa_{\rm WZ}\ge0~~;\qquad \\
\slabel{eq:ap13b}
&& \frac{(\alpha_{\rm O})_{\rm XY}}{(\alpha_{\rm O})_{\rm WZ}}\leqgr1~~;
\qquad
\frac{\kappa_{\rm XY}}{\kappa_{\rm WZ}}\geqle1~~;\quad-1\leq\kappa_{\rm WZ}
\leq0~~;                            
\label{seq:ap13}
\end{leftsubeqnarray}
which have to be read taking into consideration both lower (first alternative)
and upper (second alternative) inequalities, where the equality (third
alternative) separates the two above.
The limit of
dominant inflow/outflow rate, $\vert\kappa\vert\gg1$, for both XY and WZ
subsamples, implies
$(\alpha_{\rm O})_{\rm XY}/(\alpha_{\rm O})_{\rm WZ}\approx\kappa_{\rm XY}/
\kappa_{\rm WZ}$ to a good extent.

Then the knowledge of oxygen TDAD fractional slopes,
$s_{\rm O}=(\alpha_{\rm O})_{\rm XY}/(\alpha_{\rm O})_{\rm WZ}$, quantifies
restrictions on the flow parameter ratio, $\kappa_{\rm XY}/\kappa_{\rm WZ}$.
Oxygen TDAD fractional slopes
are listed in Table \ref{t:fras} for different inflow regimes (IR) and
different subsamples (XY, WZ).
\begin{table*}
\caption{Oxygen regression line slope ratio,
$s_{\rm O}=(\alpha_{\rm O})_{\rm XY}/(\alpha_{\rm O})_{\rm WZ}$, and flow
parameters, $\kappa_{\rm WZ}$, inferred from Eq.\,(\ref{eq:ap12}) for a
reference value, $\kappa_{\rm HH}=10$, with regard to
different subsamples, XY, WZ; XY = ND, HH; WZ = KD, ND, DD; and different
inflow regime (IR), strong inflow (SI), steady state inflow (SS), weak inflow
or outflow (WI).   The reference value assumed for the flow parameter is
valid only in WI regime.    See text for further details.}
\label{t:fras}
\begin{center}
\begin{tabular}{lllll} \hline
\multicolumn{1}{c}{IR} &
\multicolumn{1}{c}{$s_{\rm O}$} &
\multicolumn{1}{c}{$\kappa_{\rm WZ}$} &
\multicolumn{1}{c}{XY} &
\multicolumn{1}{c}{WZ} \\
\hline
SI & 0.7313 &        & ND & KD \\
SS & 1.0399 &        & ND & KD \\
WI & 1.2409 &        & ND & KD \\
   & 1.5636 & 6.0350 & HH & KD \\
   & 1.2600 & 7.7302 & HH & ND \\
   & 1.2191 & 8.0230 & HH & DD \\
\hline
\end{tabular}                     
\end{center}                      
\end{table*}                       
The results presented in Tables \ref{t:rego} and \ref{t:fras} disclose that
oxygen regression line slope ratios are consistent with unity within
$\mp\hat\sigma_{\hat\alpha}$ for disk populations and within
$\mp2\hat\sigma_{\hat\alpha}$ for disk and halo populations.

Accordingly,
inflow rate of comparable extent took place within the thick and the thin
disk during SI inflow regime $(\kappa\ll-1)$, with a difference not exceeding
a factor of about 1.3.   The same holds for SS inflow regime, in that a null
slope of the regression line relates to $\kappa=-1$ (Caimmi 2011a) and the
slopes considered are slightly larger than zero, which implies flow parameters
slightly lower than negative unity, as shown above.

Concerning WI inflow regime, outflow rate could
be conceived as strong for halo population, with a reference flow parameter
value,
$\kappa_{\rm HH}=10$ (e.g., Hartwick 1976), which via Eq.\,(\ref{eq:ap12})
yields lower values for the disk population, but within the same order of
magnitude, as shown in Table \ref{t:fras}.
On the other hand, substantial gas outflow from the disk could be
avoided if (i) the HH subsamble is representative of the inner halo and (ii)
strong outflow rate relates to the outer halo, which exhibits different trends
with respect to the inner halo in both kinematics and chemical composition
(e.g., Carollo et al. 2007, 2010).
The above considerations hold within the framework of simple MCBR chemical
evolution models, which imply (among others) the assumption of instantaneous
mixing.

In the special case where sample stars belong to a single generation, when all
stars have a similar abundance for O and Fe, an EDAD different to a delta
function could only be due to cosmic scatter.
    This is, of course,
excluded for the thin disk, where star formation is currently going on, but
should be considered for both the thick disk and the halo, which host old
populations only.

If the cosmic scatter obeys a lognormal distribution, where the mean and the
variance can be evaluated from the data, the TDAD reads (C13a; C13b):
\begin{lefteqnarray}
\label{eq:pscs}
&& (\psi)_{\rm cs}=\log\left\{\frac1{\ln10}\frac1{\sqrt{2\pi}\sigma_{\rm Q}}
\exp\left[-\frac{(\log\phi_{\rm Q}-\overline{\log\phi_{\rm Q}})^2}{2\sigma_
{\rm Q}^2}\right]\frac1{\phi_{\rm Q}}\right\}~~;
\end{lefteqnarray}
where the index, cs, denotes cosmic scatter, 
$\overline{\log\phi_{\rm Q}}=\overline{{\rm [Q/H]}}$ and 
$\sigma_{\rm Q}=\sigma_{\rm [Q/H]}$.    Related
curves, expressed by Eq.\,(\ref{eq:pscs}), are plotted in
Figs.\,\ref{f:eohk}-\ref{f:efehk} as full curves, for Q = O, Fe,
with regard to ND, KD, DD, HH subsamples, from top left
in clockwise sense.    To this aim, the values listed in Table \ref{t:parga}
have been used.   An inspection of Figs.\,\ref{f:eohk}-\ref{f:efehk}
shows both oxygen and iron EDAD cannot be due to cosmic scatter for
disk population, as expected for the thin disk, while it remains a viable 
alternative for halo population
provided HH subsample can be considered as representative.

\subsubsection{Evolution of the thick disk}
\label{evkd}

According to recent investigations (Haywood et al. 2013; Snaith et al. 2014),
the history of the thick disk was characterized by high star formation
efficiency, yielding a global mass comparable to the amount of the thin disk.
In particular, the inferred star formation rate (Snaith et al. 2014, Fig.\,2b)
relates to three different stages which, by analogy with three different
stages related to oxygen EDAD, can be similarly defined here as: A
$(13.5\appgeq t_{\rm lb}/{\rm Gyr}\appgeq12.5)$; F
$(12.5\appgeq t_{\rm lb}/{\rm Gyr}\appgeq9.5)$; E
$(9.5\appgeq t_{\rm lb}/{\rm Gyr}\appgeq8.0)$; where $t_{\rm lb}$ is the
lookback time and star formation within the thick disk is assumed to end at
$t_{\rm lb}/{\rm Gyr}\approx8.0$.

Within the framework of simple MCBR models, the transition between different
stages takes place at $\phi_{\rm Fe}\approx0.2$ and $\phi_{\rm Fe}\approx0.5$
(Fig.\,\ref{f:efeo4}, top right panel) or [Fe/H$]\approx-0.70$ and 
[Fe/H$]\approx-0.30$, respectively.   By use of the inferred [Si/Fe]-[Fe/H]
relation (Snaith et al. 2014, Fig.\,2d), the above mentioned values relate to
[Si/Fe$]\approx0.22$ and [Si/Fe$]\approx0.15$, respectively which, in turn,
via the inferred [Si/Fe]-age relation (Snaith et al. 2014, Fig.\,2a)
correspond to $t_{\rm lb}/{\rm Gyr}\approx12.5$ and
$t_{\rm lb}/{\rm Gyr}\approx10.0$, respectively.

The chemical evolution of the thick disk can be inferred from oxygen EDAD via
simple MCBR models.   The results are listed in Table
\ref{t:ra13kkz}, with regard to the normalization constant for matching
oxygen TDAD with related EDAD, $(C_{\rm U})_{\rm N}$, the flow parameter,
$\kappa_{\rm U}$, the active (i.e. available
for star formation) gas mass fraction, $(\mu_{\rm U})_{\rm f}$, the star mass
fraction, $(s_{\rm U})_{\rm f}$, the inflowed or outflowed gas mass fraction,
$(D_{\rm U})_{\rm f}$, where mass fractions are related to the initial total
mass, assumed to be entirely gaseous, $\mu_{\rm i}=1$, and the index, f, marks
the end of each stage, U = A, F, E.   For further details, an interested
reader is addressed to earlier attempts where MCBR models are formulated
(Caimmi 2011a, 2012a).
\begin{table}
\caption{Normalization constant, $(C_{\rm U})_{\rm N}$, flow parameter,
$\kappa_{\rm U}$, active gas mass fraction, $(\mu_{\rm U})_{\rm f}$, star mass
fraction, $(s_{\rm U})_{\rm f}$, inflowed or outflowed gas mass fraction,
$(D_{\rm U})_{\rm f}$, inferred from
simple MCBR models related to the thick disk
(Fig.\,\ref{f:eohk}, top right panel).
See text for further details.}
\label{t:ra13kkz}
\begin{center}
\begin{tabular}{llllll} \hline
\multicolumn{1}{l}{U} &
\multicolumn{1}{c}{$(C_{\rm U})_{\rm N}$} &
\multicolumn{1}{c}{$\kappa_{\rm U}$} &
\multicolumn{1}{c}{$(\mu_{\rm U})_{\rm f}$} &
\multicolumn{1}{c}{$(s_{\rm U})_{\rm f}$} &
\multicolumn{1}{c}{$(D_{\rm U})_{\rm f}$} \\
\hline

  &             &                &             &             &                \\
A & 1.1974E$-$1 & $-$8.7722E$-$0 & 6.9409E$-$0 & 7.6437E$-$1 & $-$6.7052E$+$0 \\
F & 5.5650E$-$1 & $-$1.4726E$-$0 & 8.2586E$-$0 & 3.5525E$-$0 & $-$1.0811E$+$1 \\
E & 1.0004E$-$0 & $+$1.8573E$-$0 & 1.6196E$-$1 & 6.3862E$-$0 & $-$5.5481E$+$0 \\
\hline                            
\end{tabular}                     
\end{center}                      
\end{table}                       

An inspection of Table \ref{t:ra13kkz} shows a net gas inflow larger than the
initial mass by a factor of about 10.8, and a final star mass fraction of
about 6.39, slightly larger than a final active + outflowed gas mass fraction
of about 5.71.   Accordingly, the thick and the thin disk are comparable in
mass provided the thin disk was built up from the gas left after thick disk
formation, as suggested in recent investigations (Haywood et al. 2013; Snaith
et al. 2014).   In addition, the flow parameter, $\kappa_{\rm U}$, equals (in
absolute value) about 8.77 in strong inflow regime, U = A; about 1.47 in
(nearly) steady state inflow regime, U = F; about 1.86 in outflow regime,
U = E.

Within the framework of simple MCBR models yielding the results listed in
Table \ref{t:ra13kkz}, the history of the thick disk can be inferred from the
above mentioned findings (Haywood et al. 2013; Snaith et al. 2014) as follows.
\begin{description}
\item[(1)\hspace{2.0mm}]
A short (about 1.0 Gyr) stage in strong inflow $(\kappa_{\rm A}\approx-8.77)$
regime, where the active gas and the star mass fraction (with respect to the
initial mass) grow up to $(\mu_{\rm A})_{\rm f}\approx6.94$ and
$(s_{\rm A})_{\rm f}\approx0.76$, respectively.
\item[(2)\hspace{2.0mm}]
A long (about 2.5 Gyr) stage in (nearly) steady state inflow
$(\kappa_{\rm A}\approx-1.47)$
regime, where the active gas and the star mass fraction 
grow up to $(\mu_{\rm F})_{\rm f}\approx8.26$ and
$(s_{\rm F})_{\rm f}\approx3.55$, respectively.
\item[(3)\hspace{2.0mm}]
A medium (about 1.5 Gyr) stage in outflow $(\kappa_{\rm A}\approx1.86)$
regime, where the active gas mass fraction decreases down to
$(\mu_{\rm E})_{\rm f}\approx0.16$ and the star mass fraction grows up to
$(s_{\rm E})_{\rm f}\approx6.39$, respectively.
\end{description}

If the global gas mass fraction left at the end of thick disk evolution,
$(\mu_{\rm E})_{\rm f}+(D_{\rm E})_{\rm f}\approx5.71$, is used for building
up the thin disk, assuming $(M_{\rm thin})_{\rm f}=5.71\,10^{10}m_\odot$ for
simplicity, then the initial thick disk mass amounts to
$(M_{\rm thick})_{\rm i}=10^{10}m_\odot$, which grows up to
$(M_{\rm thick})_{\rm f}=6.39\,10^{10}m_\odot$ at the end of evolution,
comparable to thin disk mass, as suggested in recent attempts (Haywood et al.
2013; Snaith et al. 2014).\subsection{Yields and yield ratios} \label{yiyira}

\subsubsection{Iron production via sp and np processes} \label{spnpFe}

Let Q be a selected element for which the empirical [Q/H]-[O/H] relation has
been linearly fitted according to Eq.\,(\ref{eq:QO}).
If only the contribution of sp processes is taken
into consideration, Eq.\,(\ref{eq:QO}) reduces to:
\begin{equation}
\label{eq:QspO}
[{\rm Q_{sp}/H}]=[{\rm O/H}]+b_{\rm Q_{sp}}~~;
\end{equation}
where oxygen may be thought of as sp element to a good extent.

By use of Eq.\,(\ref{eq:fiQH}),
after little algebra  Eq.\,(\ref{eq:QspO}) translates into:
\begin{lefteqnarray}
\label{eq:Qspl}
&& \phi_{\rm Q_{sp}}=B_{\rm Q_{sp}}\phi_{\rm O}~~;\qquad
B_{\rm Q_{sp}}=\exp_{10}(b_{\rm Q_{sp}})~~;
\end{lefteqnarray}
which, together with Eq.\,(\ref{eq:QOpl}), implies the following:
\begin{lefteqnarray}
\label{eq:spQ}
&& \frac{\phi_{\rm Q_{sp}}}{\phi_{\rm Q}}=\frac{B_{\rm Q_{sp}}}{B_{\rm Q}}
\left(\frac{\phi_{\rm O}}{\phi_{\rm H}}\right)^{1-a_{\rm Q}}~~; \\
\label{eq:npQ}
&& \frac{\phi_{\rm Q_{np}}}{\phi_{\rm Q}}=1-\frac{\phi_{\rm Q_{sp}}}
{\phi_{\rm Q}}~~;
\end{lefteqnarray}
where $\phi_{\rm Q_{np}}$ is the amount of the element, Q, produced via
np processes.

The particularization of Eq.\,(\ref{eq:spQ}) to the minimum abundance
exhibited by subsample stars, denoted by the index, $m$, yields:
\begin{lefteqnarray}
\label{eq:spQi}
&& \left(\frac{\phi_{\rm Q_{sp}}}{\phi_{\rm Q}}\right)_m=
\frac{B_{\rm Q_{sp}}}{B_{\rm Q}}
\left[\left(\frac{\phi_{\rm O}}{\phi_{\rm H}}\right)_m\right]^{1-a_{\rm Q}}~~;
\end{lefteqnarray}
where both $(\phi_{\rm Q_{sp}})_m$ and $B_{\rm Q_{sp}}$ are unknown while the
remaining quantities can be inferred from the data.   For sufficiently old and
low-metallicity stars, as it is in the case under consideration, the abundance
fraction, $(\phi_{\rm Q_{sp}}/\phi_{\rm Q})_m$, may safely be put equal to
unity.   Accordingly, Eq.\,(\ref{eq:spQi}) reduces to:
\begin{lefteqnarray}
\label{eq:BspB}
&& \frac{B_{\rm Q_{sp}}}{B_{\rm Q}}=
\left[\left(\frac{\phi_{\rm O}}{\phi_{\rm H}}\right)_m\right]^{a_{\rm Q}-1}~~;
\end{lefteqnarray}
and the substitution of Eq.\,(\ref{eq:BspB}) into (\ref{eq:spQ}) after some
algebra yields:
\begin{lefteqnarray}
\label{eq:fspf}
&& \frac{\phi_{\rm Q_{sp}}}{\phi_{\rm Q}}=\left[\frac
{\phi_{\rm O}/(\phi_{\rm O})_m}{\phi_{\rm H}/(\phi_{\rm H})_m}\right]^
{1-a_{\rm Q}}~~;
\end{lefteqnarray}
finally, the substitution of Eq.\,(\ref{eq:BspB}) into (\ref{eq:Qspl})
produces:
\begin{lefteqnarray}
\label{eq:fsp}
&& \phi_{\rm Q_{sp}}=B_{\rm Q}\left[\left(\frac{\phi_{\rm O}}{\phi_{\rm H}}
\right)_m\right]^{a_{\rm Q}-1}\phi_{\rm O}~~;
\end{lefteqnarray}
which, on the $({\sf O}\phi_{\rm O}\phi_{\rm Q})$ plane, represents a straight
line passing through the origin and the point, $[(\phi_{\rm O})_m,
(\phi_{\rm Q})_m]$.

\begin{figure}[t]  
\begin{center}      
\includegraphics[scale=0.8]{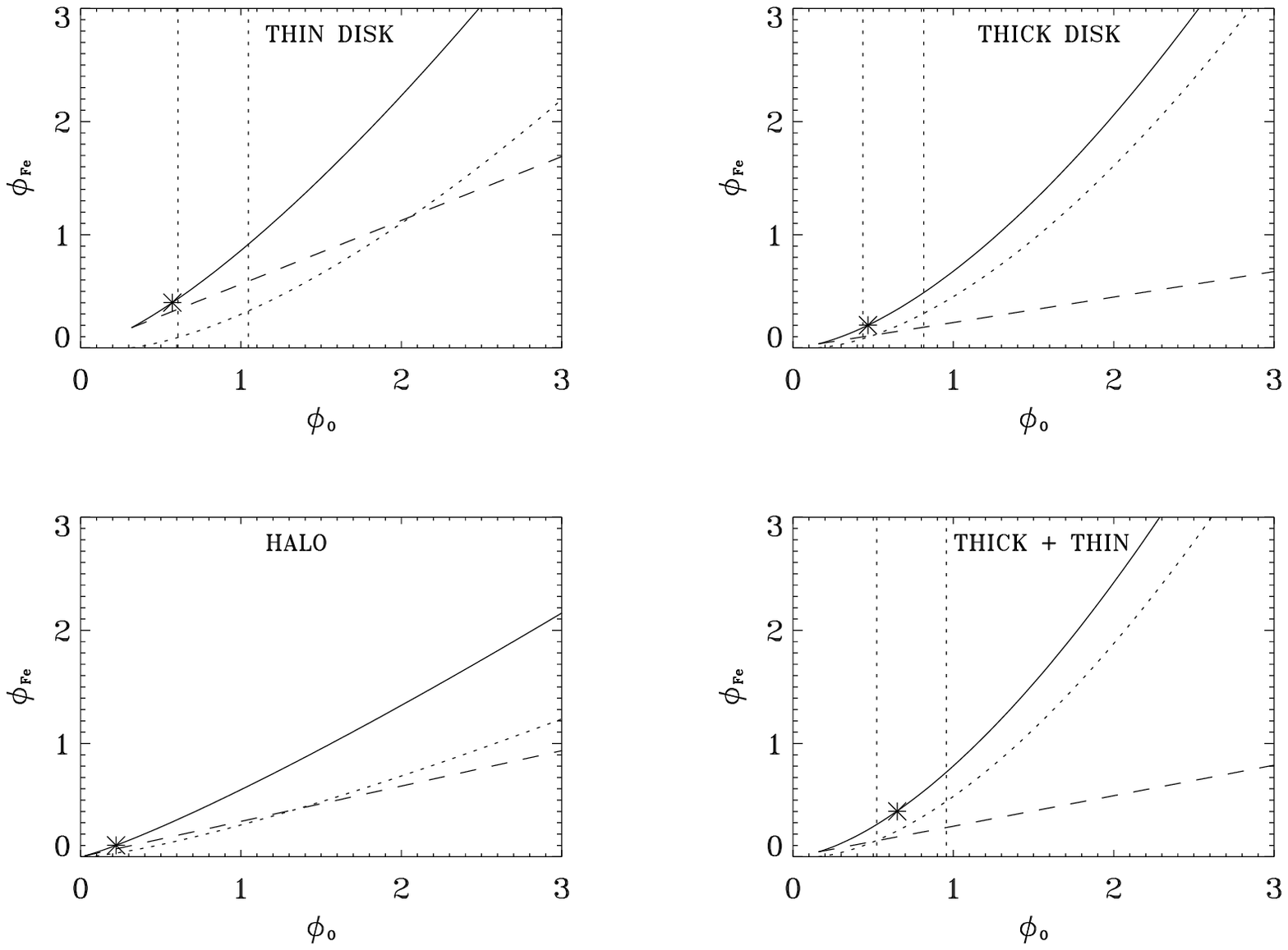}                      
\caption[ddbb]{Iron vs. oxygen mass abundance (normalized to solar values),
inferred from the empirical [Fe/H]-[O/H] relation, related to ND, KD,
DD = KD + ND + KN, HH subsamples (from top left in clockwise sense), under the
assumption of initial abundance fraction,
$(\phi_{\rm Fe\,sp}/\phi_{\rm Fe})_{\rm m}=1$.
Dashed, dotted and full curves relate to iron production via sp, np, sp+np
processes, respectively.   Dotted vertical lines mark the abundance range
related to the steady state (SS) inflow regime inferred from oxygen empirical
differential abundance
distribution (EDAD).   The abundance peak exhibited by iron EDAD is
denoted by an asterisk on related full curve.   See text for further details.}
\label{f:zofehk4}     
\end{center}       
\end{figure}

The special case of iron, related to HH, KD, ND, DD subsamples, taking
$\phi_{\rm H}=1$ (which implies hydrogen abundance in long-lived stellar
atmospheres changes only slightly with respect to the sun), is plotted in
Fig.\,\ref{f:zofehk4} for both global (full
curves) and partial (sp - dashed curves; np - dotted curves) abundances.
The abundance range related to the SS inflow regime inferred from oxygen
EDAD, plotted in Fig.\,\ref{f:eohk}, is marked by dotted vertical lines.
An asterisk on the global abundance curve denotes the peak of iron EDAD shown
in Fig.\,\ref{f:efehk}.

An inspection of Fig.\,\ref{f:zofehk4} discloses that iron production via sp
and np processes is comparable, leaving aside lower abundances, for HH
$(\phi_{\rm O}<3)$, KD $(\phi_{\rm O}<1)$, ND $(1<\phi_{\rm O}<3)$, DD
$(\phi_{\rm O}<1)$ subsample, while the contrary holds for higher abundances
exceeding the above mentioned thresholds.

The assumption of minimum subsample star abundance entirely related to sp
processes, $(\phi_{\rm Fe_{sp}})_m=(\phi_{\rm Fe})_m$, allows an explicit
expression of $B_{\rm Q_{sp}}$, Eq.\,(\ref{eq:BspB}), and then
$\phi_{\rm Q_{sp}}$, Eq.\,(\ref{eq:fspf}).   Alternatively, the following
working hypotheses can be made: (a) the above assumption, which implies the
validity of Eqs.\,(\ref{eq:BspB})-(\ref{eq:fspf}), is restriced to the thick
disk, where stars are globally older and metal poorer than in other disk
subsamples, and (b) the $\phi_{\rm Q_{sp}}-\phi_{\rm O}$ relation, expressed
by Eq.\,(\ref{eq:Qspl}), is universal i.e. $B_{\rm Q_{sp}}$ has the same value
regardless of the population, in particular it can be inferred from
Eqs.\,(\ref{eq:BspB})-(\ref{eq:fspf}) related to the thick disk.
Accordingly, dashed lines plotted in Fig.\,\ref{f:zofehk4} would be changed
into their counterpart related to the thick disk.

Then the minimum subsample (XY) star fractional abundance contributed by sp
processes,
$\left[\left(\phi_{\rm Q_{sp}}/\phi_{\rm Q}\right)_m\right]_{\rm XY}$,
can be explicitly written via the following steps.
\begin{description}
\item[(i)\hspace{2.0mm}]
Particularize
Eq.\,(\ref{eq:Qspl}) to minimum subsample star abundances and infer
$\left[\left(\phi_{\rm Q_{sp}}/\phi_{\rm Q}\right)_m\right]_{\rm XY}$.
\item[(ii)\hspace{2.0mm}]
Substitute $B_{\rm Q_{sp}}$ therein by use of Eq.\,(\ref{eq:BspB}).
\item[(iii)\hspace{2.0mm}]
Particularize Eq.\,(\ref{eq:QOpl}) to minimum subsample star abundances
and infer
$\left[\left(\phi_{\rm Q}/\phi_{\rm O}\right)_m\right]_{\rm XY}$.
\item[(iv)\hspace{2.0mm}]
Substitute
$\left[\left(\phi_{\rm Q}/\phi_{\rm O}\right)_m\right]_{\rm XY}$
into the above expression of
$\left[\left(\phi_{\rm Q_{sp}}/\phi_{\rm Q}\right)_m\right]_{\rm XY}$.
\end{description}
The result is:
\begin{lefteqnarray}
\label{eq:suni}
&& \left[\left(\frac{\phi_{\rm Q_{sp}}}{\phi_{\rm Q}}\right)_m\right]_{\rm XY}
=\frac{(B_{\rm Q})_{\rm KD}}{(B_{\rm Q})_{\rm XY}}\frac
{\left\{\left[\left(\phi_{\rm O}/\phi_{\rm H}\right)_m\right]_{\rm KD}\right\}
^{(a_{\rm Q})_{\rm KD}-1}}
{\left\{\left[\left(\phi_{\rm O}/\phi_{\rm H}\right)_m\right]_{\rm XY}\right\}
^{(a_{\rm Q})_{\rm XY}-1}}~~;\qquad
\end{lefteqnarray}
where XY = HH, ND, DD, and XY = KD yields the unit value, as expected.

In the case under consideration, Q = Fe and $\phi_{\rm H}\approx1$ may safely
be assumed, as hydrogen abundance in long-lived stellar atmospheres
is expected to remain more or less unchanged with respect to solar abundance.
Some results are listed in Table \ref{t:zegre}, where $\phi_{\rm O}$ is
directly inferred from related subsample; $\phi_{\rm Fe}$ and
$\diff\phi_{\rm Fe}/\diff\phi_{\rm O}$ are calculated via
Eq.\,(\ref{eq:QOpl});
$\phi_{\rm Fe_{sp}}/\phi_{\rm Fe}$ is calculated via Eq.\,(\ref{eq:suni});
where all values hold for both the alternatives discussed above with the
exception of the last one, which is restricted to the second alternative.
It is worth noticing comparable $(\diff\phi_{\rm Fe}/\diff\phi_{\rm O})_m$
values are shown for HH and KD subsamples, and comparable
$(\diff\phi_{\rm Fe}/\diff\phi_{\rm O})_M$
values for KD and ND subsamples, in connection with a reference maximum
oxygen abundance, $(\phi_{\rm O})_M=3$.
%
\begin{table}
\caption{Minimum subsample (sub) oxygen abundance, $(\phi_{\rm O})_m$, and
inferred iron
abundance,  $(\phi_{\rm Fe})_m$, related fractional iron abundance due to sp
processes, $(\phi_{\rm Fe_{sp}}/\phi_{\rm Fe})_m$, and slope of
$\phi_{\rm Fe}$-$\phi_{\rm O}$ relation,
$(\diff\phi_{\rm Fe}/\diff\phi_{\rm O})_m$, with the addition of its
counterpart, $(\diff\phi_{\rm Fe}/\diff\phi_{\rm O})_M$, for a reference 
maximum oxygen abundance, $(\phi_{\rm O})_M=3$.   Iron abundances and slopes
are determined via Eq.\,(\ref{eq:QOpl}).
Fractional iron abundances
are determined under the assumption of universal
$\phi_{\rm Fe_{sp}}$-$\phi_{\rm O}$ relation, according to
Eq.\,(\ref{eq:suni}), with a unit value related to the lowest
slope (KD subsample).   See text for further details.}
\label{t:zegre}
\begin{center}
\begin{tabular}{llllll} \hline
\multicolumn{1}{c}{$(\phi_{\rm O})_m$} &
\multicolumn{1}{c}{$(\phi_{\rm Fe})_m$} &
\multicolumn{1}{c}{$(\phi_{\rm Fe_{sp}}/\phi_{\rm Fe})_m$} &
\multicolumn{1}{c}{$(\diff\phi_{\rm Fe}/\diff\phi_{\rm O})_m$} &
\multicolumn{1}{c}{$(\diff\phi_{\rm Fe}/\diff\phi_{\rm O})_M$} &
\multicolumn{1}{l}{sub} \\
\hline
             &              &              &               &               &    \\
2.52855D$-$2 & 7.89204D$-$3 & 7.20182D$-$1 & 3.665260D$-$1 & 8.427390D$-$1 & HH \\
1.59541D$-$1 & 3.58617D$-$2 & 1       $ $  & 3.599973D$-$1 & 2.102907D$-$0 & KD \\
3.18326D$-$1 & 1.77952D$-$1 & 3.98951D$-$1 & 7.635178D$-$1 & 1.734610D$-$0 & ND \\
1.59541D$-$1 & 4.30047D$-$2 & 8.33903D$-$1 & 4.297271D$-$1 & 2.456852D$-$0 & DD \\
\hline       
\end{tabular}
\end{center} 
\end{table}  

\subsubsection{Iron-to-oxygen yield ratio} \label{yFeO}

With regard to Fig.\,\ref{f:efeo4},
the iron-to-oxygen yield ratio can be expressed in terms of either
iron-to-oxygen normalized abundance ratio, $\phi_{\rm Fe}/\phi_{\rm O}$, via
Eq.\,(\ref{eq:phimu}), or slope ratio of regression lines related to a
selected inflow regime for both oxygen and iron EDAD, respectively,
$(\alpha_{\rm O})_{\rm IR}/(\alpha_{\rm Fe})_{\rm IR}$, via
Eq.\,(\ref{eq:salpha}), as (C13b):
\begin{lefteqnarray}
\label{eq:fy}
&& \frac{\hat{p}_{\rm Fe}}{\hat{p}_{\rm O}}=
\frac{(Z_{\rm Fe})_\odot}{(Z_{\rm O})_\odot}
\frac{\phi_{\rm Fe}}{\phi_{\rm O}}~~; \\
\label{eq:ay}
&& \frac{\hat{p}_{\rm Fe}}{\hat{p}_{\rm O}}=
\frac{(Z_{\rm Fe})_\odot}{(Z_{\rm O})_\odot}
\frac{(\alpha_{\rm O})_{\rm IR}}{(\alpha_{\rm Fe})_{\rm IR}}~~;
\end{lefteqnarray}
where IR = SI, SS, WI, denotes the inflow regime.

In particular, SS inflow
regime relates to regression lines with slopes close to zero, as shown in
Figs.\,\ref{f:eohk} and \ref{f:efehk}, which implies slopes with different
sign for oxygen and iron, though consistent within the errors, as shown in
Tables \ref{t:rego} and \ref{t:regfe}.   On the other hand, the yield ratio
has to be non negative in the case under discussion, which implies
Eq.\,(\ref{eq:ay}) is physically meaningless or, in
other words, the yield ratio cannot be inferred via Eq.\,(\ref{eq:ay}) for SS
inflow regime.

The substitution of Eq.\,(\ref{eq:QOpl}), particularized to Q = Fe, into
Eq.\,(\ref{eq:fy}) after little algebra yields:
\begin{lefteqnarray}
\label{eq:fey}
&& \frac{\hat{p}_{\rm Fe}}{\hat{p}_{\rm O}}=
\frac{(Z_{\rm Fe})_\odot}{(Z_{\rm O})_\odot}B_{\rm Fe}
\left(\frac{\phi_{\rm O}}{\phi_{\rm H}}\right)^{a_{\rm Fe}-1}~~;
\end{lefteqnarray}
which depends on oxygen-to-hydrogen abundance ratio.   With regard to a
selected inflow regime, the combination of Eqs.\,(\ref{eq:ay}) and
(\ref{eq:fey}) produces:
\begin{lefteqnarray}
\label{eq:binf}
&& b_{\rm Fe}=\log B_{\rm Fe}=\log\left[\frac{(\alpha_{\rm O})_{\rm IR}}
{(\alpha_{\rm Fe})_{\rm IR}}\right]-(a_{\rm Fe}-1)\log
\left(\frac{\phi_{\rm O}}{\phi_{\rm H}}\right)~~;
\end{lefteqnarray}
which, keeping in mind $b_{\rm Fe}$ is inferred from the [Fe/H]-[O/H]
empirical relation, discloses the extent to which iron TDAD, plotted in
Fig.\,\ref{f:efehk}, fits to its counterpart, depending on oxygen abundance,
plotted in Fig.\,\ref{f:efeo4}.

Leaving aside F stage, where oxygen and iron regression lines show small but
opposite slopes, $(\alpha_{\rm O})_{\rm SS}>0$,
$(\alpha_{\rm Fe})_{\rm SS}<0$, and Eq.\,(\ref{eq:binf}) is undefined,
marginal disagreement, marginal agreement, satisfactory agreement are found
for KD, DD, ND and HH subsamples, respectively.   In other words, iron
TDAD plotted in Fig.\,\ref{f:efehk} is marginally inconsistent, marginally
consistent, satisfactorily consistent with the empirical [Fe/H]-[O/H] relation
for KD, DD, ND and HH subsamples, respectively, even if they fit to
related EDAD in all cases.

\begin{figure}[t]  
\begin{center}      
\includegraphics[scale=0.8]{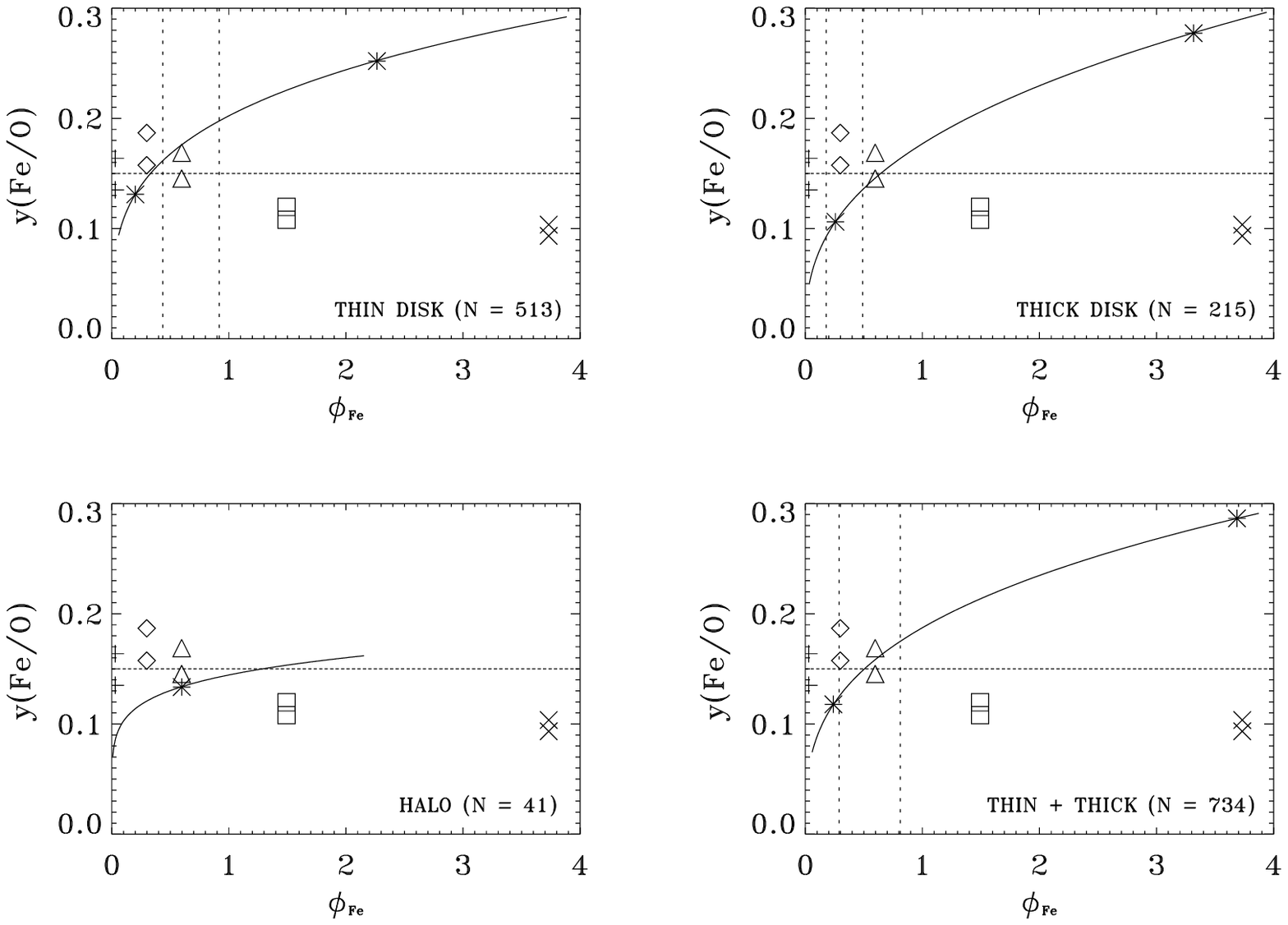}                      
\caption[ddbb]{The yield ratio, $y$(Fe/O$)=\hat{p}_{\rm Fe}/\hat{p}_{\rm O}$,
expressed by Eq.\,(\ref{eq:fey}), as a function of iron abundance,
$\phi_{\rm Fe}$, for ND, KD, DD, HH subsample (from top left in clockwise
sense).   Dotted vertical lines mark different stages inferred from oxygen
EDAD. Asterisks on each curve represent yield ratios inferred from
EDAD regression lines for both A stage (left, with the exclusion of HH
subsample) and E stage (right), regardless of iron abundance.
Theoretical yield ratios for a power-law stellar initial mass function (IMF)
with exponent, $-p=-2$ (down) and $-p=-3$ (up), restricted to
$9\le m/m_\odot\le20$, are shown for different
initial metal abundance, $Z=0.0004$ (crosses), 0.004 (diamonds),
0.008 (triangles), 0.02 (squares), 0.05 (saltires), where
$Z_{\rm Fe}/Z=(Z_{\rm Fe})_\odot/Z_\odot$ is assumed.  Power-law
exponents within the range, $-3<-p<-2$, produce results lying between the
above mentioned extreme cases.   Different mass ranges imply theoretical yield
ratios reduced by a factor up to about two.   A semiempirical yield ratio 
inferred for disk population, restricted to SNII progenitors, is shown as a
dotted horizontal line.   See text for further details.}
\label{f:yfeo4}     
\end{center}       
\end{figure}

The yield ratio, $\hat{p}_{\rm Fe}/\hat{p}_{\rm O}$, expressed by
Eq.\,(\ref{eq:fey}), as a function of iron abundance, $\phi_{\rm Fe}$, is
plotted in Fig.\,\ref{f:yfeo4} for ND, KD, DD, HH subsample (from top
left in clockwise sense), where different stages are marked by dotted vertical
lines, as in Fig.\,\ref{f:efeo4}.   Yields ratios inferred from EDAD
regression lines are represented on related curves as asterisks for both A
(left) and E (right) stage, with no connection with iron abundance.

Theoretical yield ratios for a power-law IMF
with exponent, $-p=-2$ (down) and $-p=-3$ (up), restricted to $9\le m/m_\odot
\le20$, are also shown  for different
initial metal abundance, $Z=0.0004$ (crosses), 0.004 (diamonds),
0.008 (triangles), 0.02 (squares), 0.05 (saltires), as inferred from an
earlier  research (Portinari et al. 1998) under the restriction, $Z_{\rm Fe}/Z=
(Z_{\rm Fe})_\odot/Z_\odot$, unless initial iron abundance used for the
stellar evolution is known.

Lower values, by a factor not exceeding about
two, are obtained in connection with the following mass ranges: $12\le
m/m_\odot\le20$, $9\le m/m_\odot\le30$, $12\le m/m_\odot\le30$,
$9\le m/m_\odot\le120$.   For further details on the stellar mass range
related to oxygen production and additional references, an interested reader
is addressed to a recent  research (Acharova et al. 2013) where, in particular,
an upper mass limit for SNII progenitors within the range,
$19\appleq m/m_\odot\appleq32$, with a preferred value, $m\approx23 m_\odot$,
is derived {\it a posteriori} from observations.   A reduction of iron stellar
yields by a factor of
about two has been inferred in recent  researches (e.g., Wiersma et al. 2009;
Yates et al. 2013).  Power-law exponents within the range, $-3<-p<-2$, produce
results lying between the above mentioned extreme cases.

Semiempirical oxygen-to-iron yield ratios inferred for disk population are
consistent with their counterparts inferred from star evolution theory,
$\hat{p}_{\rm Fe}/\hat{p}_{\rm O}\approx0.15$, restricted to SNII progenitors
(Acharova et al. 2013), which is shown in Fig.\,\ref{f:yfeo4} as a dotted
horizontal line.

It is worth remembering semiempirical yield ratios plotted in
Fig.\,\ref{f:yfeo4} (full curves) are expressed in the light of simple MCBR
chemical evolution
models using the regression line of the empirical [Fe/H]-[O/H] relation, via
Eq.\,(\ref{eq:fey}), where the lowest iron abundances are supposed in absence
of np processes i.e. from SNII progenitors.   Under the alternative assumption
of equal $\phi_{\rm Fe_{sp}}$-$\phi_{\rm O}$ relation for different
subsamples,
the lowest iron abundance related to sp processes should be reduced by a
factor of about 0.4 at most, according to the results listed in Table
\ref{t:zegre}.

On the other hand, theoretical yield ratios plotted in Fig.\,\ref{f:yfeo4}
(symbols) relate to SNII progenitors and, for this reason, have to be
compared with the starting point (lowest iron abundance) of each curve.
Taking into account the above mentioned uncertainties, an inspection of
Fig.\,\ref{f:yfeo4} shows agreement to an acceptable extent for subsolar
iron abundance say, while the contribution from np processes (mainly related
to SNIa progenitors) has to be considered for supersolar iron abundance (e.g.,
Wiersma et al. 2009).

If the iron-to-oxygen yield ratio increases with the metal abundance, $Z$,
while the integrated stellar yield for massive stars decreases with $Z$, it
may be interpreted as an increase of the np contribution with $Z$, which
implies an increase of SNIa rate with $Z$ compared to SNII.   

In this view, passing from lower to larger iron
abundances, semiempirical iron-to-oxygen yield ratios are increased by a
factor of about 2 for HH subsample, 2.5 for ND and DD subsample, 5
for KD subsample.   Then np processes appear to be more efficient during
thick disk evolution than during thin disk and halo evolution, according to
recent results implying a thick disk formation initially in starburst and then
more quiescent, over a time scale of 4-5 Gyr (Haywood et al. 2031; Snaith et
al. 2014).

A lower contribution to Fe abundance from np processes, by a factor
1.15, inferred {\it a posteriori} from observations in a recent research
(Acharova et al. 2013) seems to be in contradiction with the empirical
[Fe/H]-[O/H] relation, expressed by Eq.\,(\ref{eq:FeO23}) or
Eq.\,(\ref{eq:FeO34}).

\section{Conclusion}
\label{conc}

A linear [Fe/H]-[O/H] relation has been inferred from different populations
sampled in a recent  research (Ra13), namely HH (halo, $N=44$); KD
(thick disk, $N=237$); ND (thin disk, $N=538$).   Oxygen and iron
empirical differential abundance distribution (EDAD) have been determined
for different subsamples, together with related theoretical differential
abundance distribution (TDAD), within the framework of simple multistage
closed-(box+reservoir) (MCBR) chemical evolution models.

The evolution of iron vs. oxygen mass
abundance has been deduced from the empirical [Fe/H]-[O/H] relation, and iron
production via processes related to simple primary (sp) and non simple primary
(np) elements (sp and np processes, respectively), has been
estimated in two different alternatives.   Iron TDAD, inferred from empirical
[Fe/H]-[O/H] relation and oxygen EDAD, has been determined for different
subsamples.

Iron-to-oxygen yield ratios have been deduced
from the data in the framework of simple MCBR chemical evolution models,
including an example of
comparison with theoretical counterparts computed for SNII progenitors
with both subsolar and supersolar initial metallicity, under the assumption of
power-law stellar initial mass function (IMF).

Oxygen and iron TDAD have been inferred
from the data for different populations, in the opposite limit of
inhomogeneous mixing due to cosmic scatter obeying a lognormal distribution
whose mean and variance have been evaluated from the related subsample.

The main results, concerning (a) iron production via sp and np processes;
(b) evolution and flow rate; (c) iron-to-oxygen yield ratios; are summarized
as follows.
\begin{description}
\item[(a1)\hspace{2.0mm}]
Earlier results inferred from poorer subsamples (C13b) are supported in the
sense that stars display along a ``main sequence'', expressed as [Fe/H] =
$a$[O/H$]+b\mp\Delta b$.   For unit slopes, $a=1$, a main sequence relates to
constant [O/Fe] abundance ratio.   The special cases, $(a,b,\Delta b)=(3/2,
-1/8,3/8),(4/3,-2/15,4/15)$, imply only a few stars lie outside the main
sequence within the errors.
\item[(a2)\hspace{2.0mm}]
Regardless of the population, regression line slope estimators do not fit to
the unit slope within $\mp3\hat\sigma_{\hat a}$ and exhibit different values,
which implies SNIa events
contributed to a fraction of iron production to a different extent: lower for
halo population and larger for thick disk population, with thin disk
population lying between the above mentioned two.
\item[(a3)\hspace{2.0mm}]
If lowest iron abundances in subsample stars arose from sp processes, then a
comparable amount of iron was produced via sp and np processes with regard to
F stage for disk population and after iron peak for halo population.   The
above result holds for thin disk population with regard to E stage,
contrary to thick disk population, where iron production via np processes is
dominant.   A similar trend is shown by thin disk population, but not by halo
population, under the alternative assumption of equal
$\phi_{\rm Fe_{sp}}$-$\phi_{\rm O}$ relation for different subsamples leaving
its counterpart, related to thick disk population, unchanged.
\item[(a4)\hspace{2.0mm}]
Iron TDAD, inferred from
the empirical [Fe/H]-[O/H] relation and oxygen TDAD determined from linear
fits, yields a satisfactory agreement with iron EDAD via a nearly linear
trend for each stage.
\item[(a5)\hspace{2.0mm}]
A cosmic scatter, obeying a lognormal distribution, due to homogeneous mixing
in chemical evolution models, gives a TDAD which is not in agreement with the
EDAD of the disk, and therefore this is not a valid alternative to explain it.
However, this possibility remains still open for the halo.
\item[(b1)\hspace{2.0mm}]
Earlier results inferred from poorer subsamples (Caimmi and Milanese 2009;
Caimmi 2012a) are supported in the sense that oxygen EDAD can be linearly
fitted with regard to three different
stages related to different abundance ranges, namely: assembling (A, low
abundance); formation (F, middle abundance); evolution (E, high
abundance).   A similar trend is shown by iron EDAD.
Within the framework of simple MCBR chemical evolution models (Caimmi
2011a; 2012a), each stage is characterized by a different inflow regime: A -
strong inflow; F - (nearly) steady state inflow; E - weak inflow or outflow.
A (nearly) steady state inflow regime is in agreement with results from
hydrodynamical simulations (e.g., Finlator and
Dav\'e 2008; Dav\'e et al. 2011a,b, 2012). 
The outflow rate related to thick and thin disk evolution is less
then, but comparable to, outflow rate related to halo evolution, which is
known to be high (e.g., Hartwick 1976).   A low outflow rate related to disk
evolution would imply (i) the halo subsample considered in the current research
is representative of the inner halo, and (ii) a higher outflow rate relates to
the outer halo.
\item[(b2)\hspace{2.0mm}]
Under the assumption that the gas lef after the evolution of
the thick disk was used for building up the thin disk, the two subsystems
exhibit comparable masses according to recent investigations (Haywood et al.
2013; Snaith et al. 2014).
\item[(c1)\hspace{2.0mm}]
Within the framework of simple MCBR chemical evolution models (Caimmi
2011a; 2012a), iron-to-oxygen yield ratios, inferred from the empirical
[Fe/H]-[O/H] relation for different subsamples, are consistent with
theoretical results from SNII progenitor nucleosynthesis (Portinari et al.
1998; Wiersma et al. 2009), provided substantial iron production arises from
SNIa events for supersolar abundances.
\end{description}



\begin{thebibliography}{}
%

\bibitem{} Acharova, I.A., Gibson, B.K., Mishurov, Yu.N., Kovtyukh, V.V.:
           2013, Astron. Astrophys., 557, A107.

\bibitem{} Bekki, K.: 2013, Astrophys. J., 779, 9. 

\bibitem{} Caimmi, R.: 2007, New Astron., 12, 289.


\bibitem{} Caimmi, R.: 2011a, Ser. Astron. J., 183, 37.

\bibitem{} Caimmi, R.: 2011b, New Astron., 16, 337.

\bibitem{} Caimmi, R.: 2012a, Ser. Astron. J., 185, 35.

\bibitem{} Caimmi, R.: 2012b,  Intellectual Archive, 1, 71, ISSN 1929-4700
                       Toronto. (arxiv 1111.2680).

\bibitem{} Caimmi, R.: 2013a, Ser. Astron. J., 186, 25. (C13a).

\bibitem{} Caimmi, R.: 2013b, Ser. Astron. J., 187, 19. (C13b).

\bibitem{} Caimmi, R., Milanese, E.: 2009, Astrophys. Space Sci., 323, 147.


\bibitem{} Carollo, D., Beers, T.C., Lee, Y.S., et al.: 2007, Nature, 318,
           1020.

\bibitem{} Carollo, D., Beers, T.C., Chiba, M., et al.: 2010, Astrophys. J.,
           712, 692.


\bibitem{} Carretta, E., Gratton, R.G.,  Sneden, C.: 2000, Astron. Astrophys.,
           356, 238.


\bibitem{} Conroy, C., Dutton, A.A., Graves, G.J., Trevor Mendel, J., van
           Dokkum, P.G.: 2013, Astrophys. J., 776, L26.


\bibitem{} Dav\'e, R., Finlator, K., Oppenheimer, B.D.: 2011a, Mon. Not. R.
           Astron. Soc., 415, 11.

\bibitem{} Dav\'e, R., Finlator, K., Oppenheimer, B.D.: 2011b, Mon. Not. R.
           Astron. Soc., 416, 1354.

\bibitem{} Dav\'e, R., Finlator, K., Oppenheimer, B.D.: 2012, Mon. Not. R.
           Astron. Soc., 421, 98.


\bibitem{} Doane, J.S., Mathews, W.G.: 1993, Astrophys. J., 419, 573.


\bibitem{} Finlator, K., Dav\'e, R.: 2008, Mon. Not. R. Astron. Soc., 385,
           2181.


\bibitem{} Gratton, R.G., Carretta, E., Matteucci, F., Sneden, C.: 2000,
           Astron. Astrophys., 358, 671.


\bibitem{} Hartwick, F.D.A.: 1976, Astrophys. J., 209, 418.

\bibitem{} Haywood, M.: 2001, Mon. Not. R. Astron. Soc., 325, 1365.


\bibitem{} Haywood, M., Di Matteo, P., Lehnert, M.D., Katz, D., G\'omez, A.:
           2013, Astron. Astrophys., 560, A109.


\bibitem{} Isobe, T., Feigelson, E.D., Akritas, M.G., Babu, G.J.: 1990,
           Astrophys. J., 364, 104.


\bibitem{} Israelian, G., Rebolo, R., Garcia-Lopez, R.J., Bonifacio, P.,
           Molaro, P., Basri, G., Shchukina, N.: 2001, Astrophys. J., 551,
           833.


\bibitem{} Lopez-Sanchez, A.R.: 2010, Astron. Astrophys., 521, A63.


\bibitem{} Malinie, G., Hartmann, D.H.,
           Clayton, D.D., Mathews, G.J.: 1993, Astrophys. J,. 413, 633.


\bibitem{} Pagel, B.E.J.: 1989, The G-dwarf Problem and Radio-active
           Cosmochronology. In: Beckman J.E., Pagel B.E.J. (eds.)
           Evolutionary Phenomena in Galaxies, p.\,201. Cambridge
           University Press, Cambridge.

\bibitem{} Pagel, B. E. J., Tautvaisiene, G.: 1995, Mon. Not. R. Astron. Soc.,
           276, 505.


\bibitem{} Pagel, B.E.J., Patchett, B.E.: 1975, Mon. Not. R. Astron. Soc.,
           172, 13.


\bibitem{} Portinari, L., Chiosi, C., Bressan, A.: 1998, Astron. Astrophys.,
           334, 505.


\bibitem{} Ram{\'\i}rez, I., Allende Prieto, C., Lambert, D.L.: 2007, Astron.
           Astrophys., 757, 164.

\bibitem{} Ram{\'\i}rez, I., Mel\'endez, J., Chanam\'e, J.: 2012, Astrophys.
           J., 757, 164.  

\bibitem{} Ram{\'\i}rez, I., Allende Prieto, C., Lambert, D.L.: 2013,
           Astrophys. J., 764, 78.  (Ra13).


\bibitem{} Rieke, G.H., Loken, K., Rieke, M.J., Tamblyn, P.: 1993, Astrophys.
          J., 412, 99.

\bibitem{} Rocha-Pinto, H.J., Maciel, W.J.: 1996, Mon. Not. R. Astron. Soc.,
           279, 447.


\bibitem{} Ryan, S.G., Norris, J.E.: 1991, Astron. J., 101, 1865. 


\bibitem{} Searle, L., 1972. Star Formation and the Chemical History of
           Galaxies.   In: Cayrel de Strobel, G., Delplace, A.M. (eds.)
           L'Ages des Etoiles, Observatoire de Paris-Meudon, p.\,52.

\bibitem{} Searle, L., Sargent, W.L.W.: 1972, Astrophys. J., 173, 25.


\bibitem{}  Snaith, O.N., Haywood, M., Di Matteo, P., Lehnert, M.D.: 2014,
            Astrophys. J., 781, L31.


\bibitem{}  Takada-Hidai, M., Takeda, Y., Sato, S., Sargent, W.L., Lu, L.,
            Barlow, T., Jugeku, J.: 2001, New Astron. Rev., 45, 549.


\bibitem{} Wiersma, R.P.C., Schaye, J., Theuns, T., Dalla Vecchia, C.,
           Tornatore, L.: 2009, Mon. Not. R. Astron. Soc., 399, 574.

\bibitem{} Wilmes, M., K\"oppen, J.: 1995, Astron. Astrophys., 294, 47.

\bibitem{}  Woosley, S. E., Weaver, T. A., 1995. Astrophys. J. Supp. 101, 181.


\bibitem{} Yates, R.M., Henriques, B., Thomas, P.A., Kauffmann, G., Johansson,
           J., White, S.D.M.: 2013, Mon. Not. R. Astron. Soc., 435, 3500.

\end{thebibliography}
\end{document}